\documentclass[longauth]{aa}

\usepackage[switch]{lineno}
\linenumbers
\usepackage{graphicx}
\usepackage{arydshln}
\usepackage{subcaption}
\usepackage{hyperref}
\hypersetup{colorlinks=true,linkcolor=blue,filecolor=blue,urlcolor=blue,citecolor=blue}
\usepackage{appendix}
\usepackage{comment}
\usepackage{anyfontsize}
\usepackage{orcidlink}
\usepackage{ulem}
\usepackage{txfonts}
\makeatletter
\renewcommand*\aa@pageof{, page \thepage{} of \pageref*{LastPage}}
\makeatother

\begin{document} 

   \title{Characterization of Markarian 421 during its most violent year: Multiwavelength variability and correlations}
   \titlerunning{Mrk 421 during its most violent year}

    %
    \author{\normalsize K.~Abe\inst{1} \and
    S.~Abe\inst{2}\orcidlink{0000-0001-7250-3596} \and
    J.~Abhir\inst{3}\orcidlink{0000-0001-8215-4377} \and
    A.~Abhishek\inst{4}\orcidlink{0009-0005-5239-7905} \and
    V.~A.~Acciari\inst{5}\orcidlink{0000-0001-8307-2007} \and
    A.~Aguasca-Cabot\inst{6}\orcidlink{0000-0001-8816-4920} \and
    I.~Agudo\inst{7}\orcidlink{0000-0002-3777-6182} \and
    T.~Aniello\inst{8} \and
    S.~Ansoldi\inst{9,42}\orcidlink{0000-0002-5613-7693} \and
    L.~A.~Antonelli\inst{8}\orcidlink{0000-0002-5037-9034}
% ...
    }
    \institute { Japanese MAGIC Group: Department of Physics, Tokai University, Hiratsuka, 259-1292 Kanagawa, Japan
    \and Japanese MAGIC Group: Institute for Cosmic Ray Research (ICRR), The University of Tokyo, Kashiwa, 277-8582 Chiba, Japan
    \and ETH Z\"urich, CH-8093 Z\"urich, Switzerland
    \and Universit\`a di Siena and INFN Pisa, I-53100 Siena, Italy
    \and Institut de F\'isica d'Altes Energies (IFAE), The Barcelona Institute of Science and Technology (BIST), E-08193 Bellaterra (Barcelona), Spain
    \and Universitat de Barcelona, ICCUB, IEEC-UB, E-08028 Barcelona, Spain
    \and Instituto de Astrof\'isica de Andaluc\'ia-CSIC, Glorieta de la Astronom\'ia s/n, 18008, Granada, Spain
    \and National Institute for Astrophysics (INAF), I-00136 Rome, Italy
    \and Universit\`a di Udine and INFN Trieste, I-33100 Udine, Italy
    \and Max-Planck-Institut f\"ur Physik, D-85748 Garching, Germany
    \and Universit\`a di Padova and INFN, I-35131 Padova, Italy
    \and Technische Universit\"at Dortmund, D-44221 Dortmund, Germany
    \and Croatian MAGIC Group: University of Zagreb, Faculty of Electrical Engineering and Computing (FER), 10000 Zagreb, Croatia
    \and Centro Brasileiro de Pesquisas F\'isicas (CBPF), 22290-180 URCA, Rio de Janeiro (RJ), Brazil
    \and IPARCOS Institute and EMFTEL Department, Universidad Complutense de Madrid, E-28040 Madrid, Spain
    \and Instituto de Astrof\'isica de Canarias and Dpto. de  Astrof\'isica, Universidad de La Laguna, E-38200, La Laguna, Tenerife, Spain
    \and University of Lodz, Faculty of Physics and Applied Informatics, Department of Astrophysics, 90-236 Lodz, Poland
    \and Centro de Investigaciones Energ\'eticas, Medioambientales y Tecnol\'ogicas, E-28040 Madrid, Spain
    \and Departament de F\'isica, and CERES-IEEC, Universitat Aut\`onoma de Barcelona, E-08193 Bellaterra, Spain
    \and Universit\`a di Pisa and INFN Pisa, I-56126 Pisa, Italy
    \and INFN MAGIC Group: INFN Sezione di Bari and Dipartimento Interateneo di Fisica dell'Universit\`a e del Politecnico di Bari, I-70125 Bari, Italy
    \and Department for Physics and Technology, University of Bergen, Norway
    \and INFN MAGIC Group: INFN Sezione di Torino and Universit\`a degli Studi di Torino, I-10125 Torino, Italy
    \and Croatian MAGIC Group: University of Rijeka, Faculty of Physics, 51000 Rijeka, Croatia
    \and Universit\"at W\"urzburg, D-97074 W\"urzburg, Germany
    \and Japanese MAGIC Group: Physics Program, Graduate School of Advanced Science and Engineering, Hiroshima University, 739-8526 Hiroshima, Japan
    \and Deutsches Elektronen-Synchrotron (DESY), D-15738 Zeuthen, Germany
    \and Armenian MAGIC Group: ICRANet-Armenia, 0019 Yerevan, Armenia
    \and Croatian MAGIC Group: University of Split, Faculty of Electrical Engineering, Mechanical Engineering and Naval Architecture (FESB), 21000 Split, Croatia
    \and Croatian MAGIC Group: Josip Juraj Strossmayer University of Osijek, Department of Physics, 31000 Osijek, Croatia
    \and Finnish MAGIC Group: Finnish Centre for Astronomy with ESO, Department of Physics and Astronomy, University of Turku, FI-20014 Turku, Finland
    \and Saha Institute of Nuclear Physics, A CI of Homi Bhabha National Institute, Kolkata 700064, West Bengal, India
    \and Inst. for Nucl. Research and Nucl. Energy, Bulgarian Academy of Sciences, BG-1784 Sofia, Bulgaria
    \and Japanese MAGIC Group: Department of Physics, Yamagata University, Yamagata 990-8560, Japan
    \and Finnish MAGIC Group: Space Physics and Astronomy Research Unit, University of Oulu, FI-90014 Oulu, Finland
    \and Japanese MAGIC Group: Chiba University, ICEHAP, 263-8522 Chiba, Japan
    \and Japanese MAGIC Group: Institute for Space-Earth Environmental Research and Kobayashi-Maskawa Institute for the Origin of Particles and the Universe, Nagoya University, 464-6801 Nagoya, Japan
    \and Japanese MAGIC Group: Department of Physics, Kyoto University, 606-8502 Kyoto, Japan
    \and INFN MAGIC Group: INFN Roma Tor Vergata, I-00133 Roma, Italy
    \and University of Geneva, Chemin d'Ecogia 16, CH-1290 Versoix, Switzerland
    \and Japanese MAGIC Group: Department of Physics, Konan University, Kobe, Hyogo 658-8501, Japan
    \and also at International Center for Relativistic Astrophysics (ICRA), Rome, Italy
% ...
}
     
   \date{Received XX XX, 2023; accepted XX XX, 2023}

 
  \abstract
   {}
   {Mrk\,421 was in its most active state around early 2010, which led to the highest  TeV gamma-ray flux ever recorded from any active galactic nuclei (AGN). We aim to characterize the multiwavelength behavior during this exceptional year for Mrk\,421, and evaluate whether it is consistent with the picture derived with data from other less exceptional years.} 
   {We investigated the period from November 5, 2009, (MJD 55140) until July 3, 2010, (MJD 55380) with extensive coverage from very-high-energy (VHE; E$\,>\,$100$\,$GeV) gamma rays to radio with MAGIC, VERITAS, \textit{Fermi}-LAT, \textit{RXTE}, \textit{Swift}, GASP-WEBT, VLBA, and a variety of additional optical and radio telescopes. We characterized the variability by deriving fractional variabilities as well as power spectral densities (PSDs). In addition, we investigated images of the jet taken with VLBA and the correlation behavior among different energy bands.}
   {Mrk~421 was in widely different states of activity throughout the campaign, ranging from a low-emission state to its highest VHE flux ever recorded. We find the strongest variability in X-rays and VHE gamma rays, and PSDs compatible with power-law functions with indices around 1.5. We observe strong correlations between X-rays and VHE gamma rays at zero time lag with varying characteristics depending on the exact energy band. We also report a marginally significant ($\sim 3\sigma$) positive correlation between high-energy (HE; E$\,>\,$100$\,$MeV) gamma rays and the ultraviolet band. We detected marginally significant ($\sim 3\sigma$) correlations between the HE and VHE gamma rays, and between HE gamma rays and the X-ray, that disappear when the large flare in February 2010 is excluded from the correlation study, hence indicating the exceptionality of this flaring event in comparison with the rest of the campaign. The 2010 violent activity of Mrk421 also yielded the first ejection of features in the VLBA images of the jet of Mrk\,421. Yet the large uncertainties in the ejection times of these unprecedented radio features prevent us from firmly associating them to the specific flares recorded during the 2010 campaign. We also show that the collected multi-instrument data are consistent with a scenario where the emission is dominated by two regions, a compact and extended zone, which could be considered as a simplified implementation of an energy-stratified jet as suggested by recent \textit{IXPE} observations.}
   {}

   \keywords{BL Lacertae objects:  individual: Mrk 421 galaxies:  active   gamma rays:  general radiation mechanisms:  nonthermal}

   \maketitle
%


\section{Introduction} \label{sec:intro}
Blazars are a class of jetted active galactic nuclei (AGN). The relativistic plasma jet is oriented with a small angle with respect to the line of sight. Blazars emit across the full electromagnetic spectrum, ranging from radio to high-energy (HE; E$\,>\,$100$\,$MeV) and very-high-energy (VHE; E$\,>\,$100$\,$GeV) gamma rays. Blazars that show no or very faint emission lines in their optical emission are referred to as BL Lac-type objects~\citep{1995PASP..107..803U}.\par

The broadband emission of BL Lac-type objects is dominated by non-thermal radiation from the jet. The spectral energy distribution (SED) exhibits two large features in the form of two large bumps~\citep[see e.g.][]{abdo:2011}. Measurements of spectral and polarization characteristics strongly indicate that the first bump originates from synchrotron radiation produced by relativistic electrons and positrons moving in the magnetic field within the jet. The origin of the second bump is still under debate and is more difficult to determine. Possible leptonic scenarios include electron inverse Compton (IC) scattering on synchrotron photons originating from the first bump, so called synchrotron self-Compton (SSC)~\citep{1992ApJ...397L...5M, 1998MNRAS.301..451G, 1999ApJ...521..145M}, or in certain cases additionally on external target photons~\citep{1992A&A...256L..27D, 1994ApJ...421..153S}. Hadronic scenarios can also provide explanations for the gamma-ray emission~\citep{1993A&A...269...67M, 2001APh....15..121M, 2015MNRAS.448..910C}. BL Lac-type objects can be classified by their peak frequency of the synchrotron bump~\citep{1995PASP..107..803U,2017A&ARv..25....2P, 2010ApJ...716...30A}. Blazars with a peak frequency of $\nu_s<10^{14}\,$Hz are referred to as low synchrotron peaked blazars (LSPs), with a peak frequency $10^{14}\,\textrm{Hz}<\nu_s<10^{15}\,$Hz as intermediate synchrotron peaked blazars (ISPs) and with $\nu_s>10^{15}\,\textrm{Hz}$ as high synchrotron peaked blazars (HSPs).\par

Markarian 421 (Mrk$\,$421; RA=11$^\text{h}$4$^{\prime}$27.31$^{\prime\prime}$, Dec=38$^{\circ}$12$^{\prime}$31.8$^{\prime\prime}$, J2000, z=0.031) is an archetypal and very close HSP. It is among the most studied sources in the VHE sky. Mrk$\,$421 was found in states of extreme activity in the X-ray and VHE gamma-ray bands during previous campaigns~\citep[e.g.][]{1996Natur.383..319G, 2008ApJ...677..906F, 2015A&A...578A..22A, 2020ApJS..248...29A, 2020ApJ...890...97A}. Among all campaigns, 2010 stands out as the most active and violent year seen in the VHE emission from Mrk$\,$421. In February 2010, the Very Energetic Radiation Imaging Telescope Array System (VERITAS) detected the highest VHE flux recorded to date from Mrk$\,$421~\citep{2020ApJ...890...97A}. The highest flux measurement reached a level of $\sim$27 Crab Units (C.U.) above 1$\,$TeV, making it the brightest VHE gamma-ray flux ever recorded for an AGN. The high photon statistic during such flaring activity allowed for a binning of the flux in 2-minute intervals. A cross-correlation study with the optical band revealed a significant positive correlation with a time lag of $\sim$25-55 minutes. Additional correlation studies between the VHE and X-ray bands showed a complex and fast-varying correlation from linear to quadratic behavior down to no correlation at times. Shortly after, in March 2010, a decaying flare was detected by the Florian Goebel Major Atmospheric Gamma Imaging Cherenkov (MAGIC) and VERITAS telescopes~\citep{2015A&A...578A..22A}. The detailed and broad multiwavelength (MWL) coverage enabled the construction of 13 consecutive daily SEDs. The emission could be successfully modeled by a single-SSC zone. However, as described in \cite{2015A&A...578A..22A}, a better model-data agreement and a more natural explanation of the SED evolution could be achieved with a two-zone scenario. The variation of only a few model parameters could successfully describe the SED variation and made the underlying particle population a very plausible cause of the variable emission. \par
For the first time, this work exploits the full campaign covering a time period spanning from November 2009 until June 2010 with a broad MWL coverage. High-resolution radio data taken with the Very Long Baseline Interferometry (VLBI) technique add observations of the jet structure to the campaign from mid January 2010 to early July 2012. This wealth of data allows for a detailed study of the variability and correlation behavior of Mrk$\,$421 during its most violent recorded year. \par

\section{Multiwavelength light curves} \label{sec:MWL_LC}

The study involves data from several instruments covering the emission from radio frequencies to VHE in the time period from November 5, 2009, (MJD 55140) until July 3, 2010, (MJD 55380). The key instruments used are MAGIC, VERITAS, the Large Area Telescope (LAT) on board the \textit{Fermi Gamma-ray Space Telescope} (\textit{Fermi}-LAT), the X-Ray telescope (XRT) and the Ultra-violet Optical Telescope (UVOT) on board the \textit{Neil Gehrels Swift Observatory (Swift)} and the GLAST-AGILE Support Program - Whole Earth Blazar Telescope (GASP-WEBT) network of optical telescopes. Additional telescopes in the X-ray, optical and radio support these. The details of the data analysis of the individual instruments are given in the Appendix in Sec.~\ref{sec:data_analysis}. \par

\begin{figure*}[h!]
  \centering
  \includegraphics[width=0.95\textwidth]{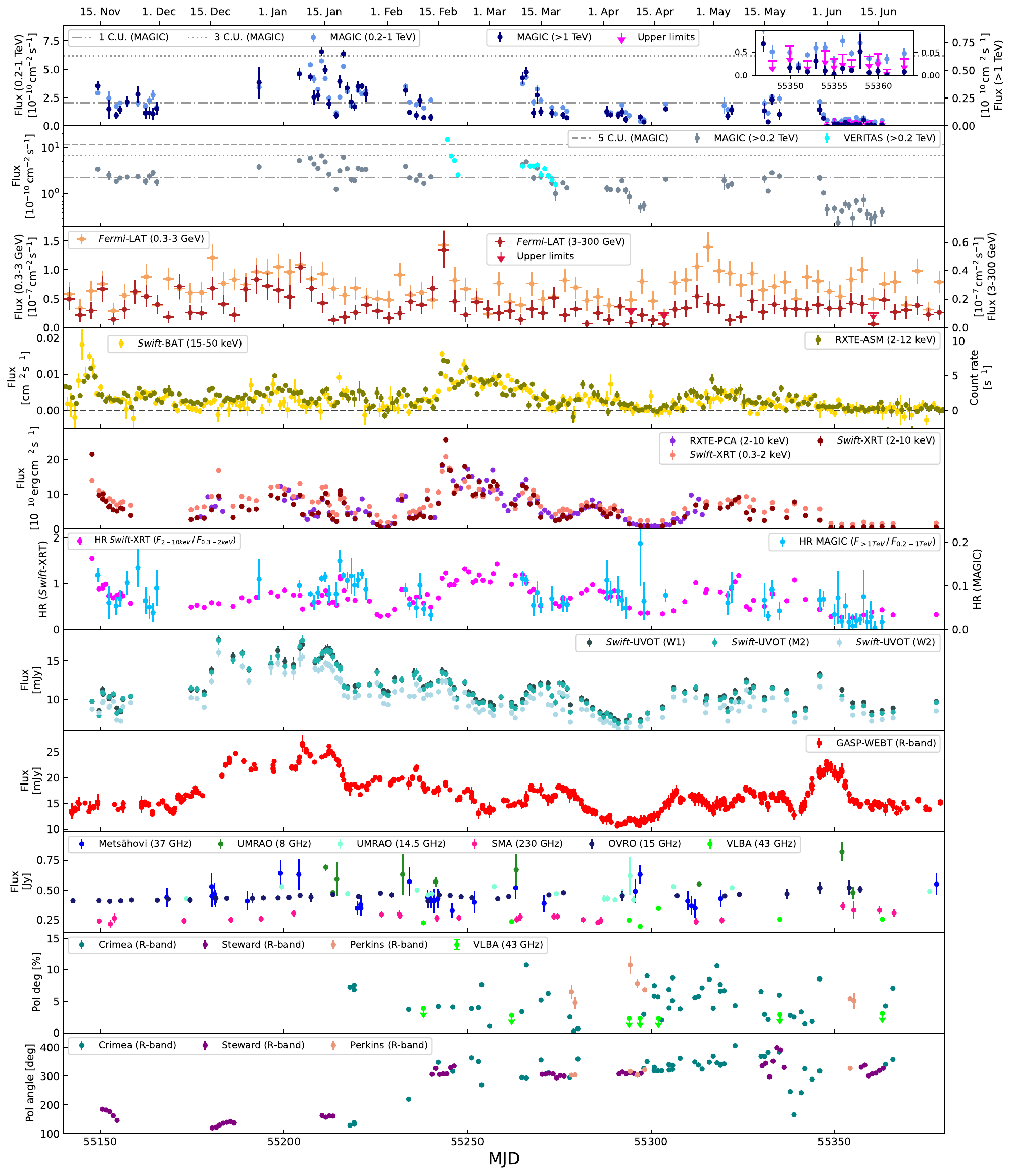}
 \caption{MWL light curves covering the time period from November 5, 2009, (MJD 55140) to July 3, 2010, (MJD 55380). Top to bottom: MAGIC fluxes in daily bins for two energy bands (note the two different y-axes); VHE fluxes obtained from MAGIC and VERITAS above 0.2\,TeV (in log scale); \textit{Fermi}-LAT fluxes in 3-day bins in two energy bands; X-ray fluxes in 1-day bins from the all-sky monitors \textit{Swift}-BAT and \textit{RXTE}-ASM; X-ray fluxes from the pointing instruments \textit{Swift}-XRT and \textit{RXTE}-PCA; hardness ratio between the high and low-energy fluxes of \textit{Swift}-XRT and between the two VHE bands of MAGIC (note the two different y-axes); optical R-band data from GASP-WEBT; radio data from Mets\"ahovi, UMRAO, SMA, OVRO and VLBA; polarization degree and polarization angle observations in the optical from the Steward and Perkins observatories and in radio from VLBA.}
 \label{fig:MWL_LC}
\end{figure*}

The top row of Fig.~\ref{fig:MWL_LC} shows the light curves provided by the MAGIC telescopes in the 0.2-1$\,$TeV and >1$\,$TeV band. For comparison, multiples of the flux of the Crab Nebula\footnote{The flux of the Crab Nebula used in this work is taken from ~\citet{aleksic:2016}} are given with dashed-dotted lines. At the beginning of the campaign in November 2009, Mrk$\,$421 was already in an elevated state of activity with an emission of around 1 C.U. in both bands. The average emission state of the source estimated by Whipple over a time span of 14 years is around 0.45 C.U.~\citep{2014APh....54....1A}. After a small gap of observations, Mrk$\,$421 showed a strong flaring activity reaching over 3 C.U. above 1$\,$TeV in January 2010. The emission showed strong variability by up to a factor of almost three on a daily timescale. We found significant intranight variability on January 15 (see Appendix~\ref{sec:appendix_intranight}). After a short phase at around 1 C.U., another observational gap due to adverse weather conditions is present in the MAGIC telescope data. Following the gap, a decaying flare starting at around 2 C.U. could be observed, which is discussed in great detail in~\citet{2015A&A...578A..22A}. Mrk$\,$421 showed average emission for the following two months until it entered an especially low emission state around June, going below 0.25 C.U.. \par 
The second row shows the combined VHE light curve provided by MAGIC and VERITAS on a logarithmic scale. Due to the additional VERITAS data, the gap in February 2010 is covered. These observations yielded the largest VHE flux observed to date in an AGN, reaching almost 10 C.U. for the daily average. Using a shorter binning, the peak flux values reached $\sim$15 C.U. above 200$\,$GeV~\citep{2020ApJ...890...97A}. \par
The third row displays the emission in high-energy gamma rays (HE; E > 100\,MeV) measured by \textit{Fermi}-LAT in a three-day binning in the 0.3\,GeV to 3\,GeV and 3\,GeV to 300\,GeV band. During the large VHE flare in February, both bands show increased emission and reached their highest flux coincident with the highest VHE flux point. The other two VHE flares during January and March do not show increased activity in HE gamma rays. We additionally checked if daily binning reveals any activity on a shorter scale during these two flares that might be hidden with a 3-day binning but did not find significant variability. \par
In the fourth and fifth rows, the results of the X-ray are shown. A short outburst is detected in November 2009, around MJD 55145. The flare faints over a few days, and the source remains at a comparably lower activity level without any noticeable outbursts. Coinciding with the VHE flare in February, all wavebands show a sharp increase in flux, which slowly decreases in the following month. At the end of the campaign, the X-ray emission also entered a remarkably low state. \par 
Using the two separate VHE and X-ray light curves, hardness ratios (i.e. the ratio of the flux in the higher-energy band to the flux in the lower-energy band) can be obtained and are shown in row six. Overall, the source showed a harder-when-brighter trend in VHE and X-ray, which has been observed frequently in the past~\citep[e.g.][]{2021MNRAS.504.1427A, 2021A&A...655A..89M}. This behavior is investigated in more detail in Appendix~\ref{sec:appendix_HR}. \par
Row seven and eight contain the UV and optical fluxes, respectively. Due to the closeness in frequency, the trends are quite similar. Both light curves reach their highest emission coincident to the January VHE flare. When the subsequent February flare starts, there is no clear visible flare in the UV or optical. \par
The ninth row shows the fluxes obtained from radio observations. The instruments cover a variety of frequencies, which makes flux comparisons challenging. Overall, the source showed a low activity and a low variability in all radio bands. \par
The second to last row reports the polarization degree, which shows some strong fluctuations throughout the campaign, going from values as low as 0.3\% up to $\sim11$\% around a mean of $\sim5$\%. The corresponding electric vector polarization angles (EVPA) in the optical are shown in the last row. The data were adjusted for the intrinsic 180$^{\circ}$ ambiguity as reported in~\citet{2017MNRAS.472.3789C}. The EVPA shows a constant behavior at the beginning at around $\sim150^{\circ}$ followed by a wide rotation happening around MJD 55230 (early February). For the remaining campaign, it fluctuates around $\sim350^{\circ}$. Since the EVPA rotates by around $200^{\circ}$ during a time interval with very limited coverage (only one measurement), one cannot completely exclude that this rotation arises from an improper correction of the 180$^{\circ}$ ambiguity. If the rotation is real, there might be a possible connection with the large flare in February at higher energies.  \par

\section{VLBA observations of the jet evolution} \label{sec:VLBA}
\begin{figure*}
\centering
  \resizebox{\hsize}{!}{\includegraphics{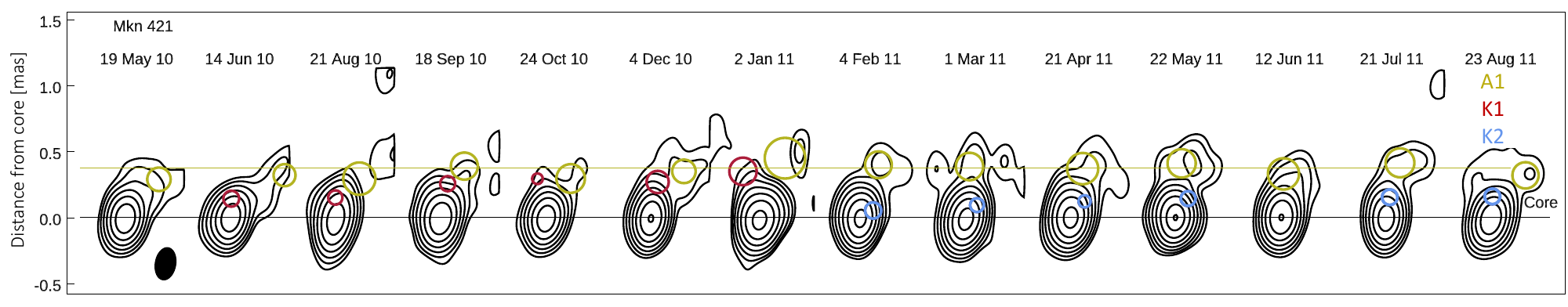}}
  \caption{Total intensity VLBA images from May 2010 to August 2011. The black and yellow lines show the average positions of the stationary features A0 and A1. The yellow, red and blue circles show the fitted positions of A1, K1, and K2, respectively.}
  \label{fig:VLBA}
\end{figure*}

\begin{center}
\begin{table*}[h]
\caption{Average parameters of the features shown in Fig.~\ref{fig:VLBA}} 
\centering
\begin{tabular}{ c c c c c c}
Component &  Number of epochs  &   Flux (Jy) & Distance (mas) & Theta (deg) & Size (FWHM,mas) \\ 
\hline \hline
A0   &  31  & $0.210\pm0.048$  &  0.0	          &	 ...           &	$0.072\pm0.030$ \\
A1   &  24  & $0.025\pm0.012$  &  $0.378\pm0.062$ &	$-28.9\pm14.3$ &	$0.267\pm0.117$	\\
K1   &   5  & $0.011\pm0.008$  &  $0.29\pm0.18$	  & $-34.2\pm27.2$ &	$0.146\pm0.131$ \\
K2   &   8  & $0.060\pm0.005$  &  $0.20\pm0.12$	  & $-37.3\pm7.6$  &	$0.145\pm0.036$ \\
\hline \hline
\end{tabular}
\label{tab:VLBA_params}
\end{table*}
\end{center}

\begin{center}
\begin{table*}[h]
\caption{Kinematic properties of the knots K1 and K2.} 
\centering
\begin{tabular}{ c c c c c c }
Knot & Proper motion (mas/yr)     & Phi (deg) & Apparent speed (c) &  Ejection time (yr) & Ejection time (MJD)  \\ 
\hline \hline
K1	& $0.775\pm0.225$ &	 $49.5\pm15.2$	& $1.56\pm0.45$  & $2009.77\pm0.24$  & $55112\pm88$	\\
K2	& $0.157\pm0.033$ &	$-21.8\pm20$	& $0.32\pm0.07$  & $2010.56\pm0.43$  & $55400\pm157$ \\
\hline \hline
\end{tabular}
\label{tab:VLBA_kin}
\end{table*}
\end{center}

\begin{figure}
\centering
  \resizebox{\hsize}{!}{\includegraphics{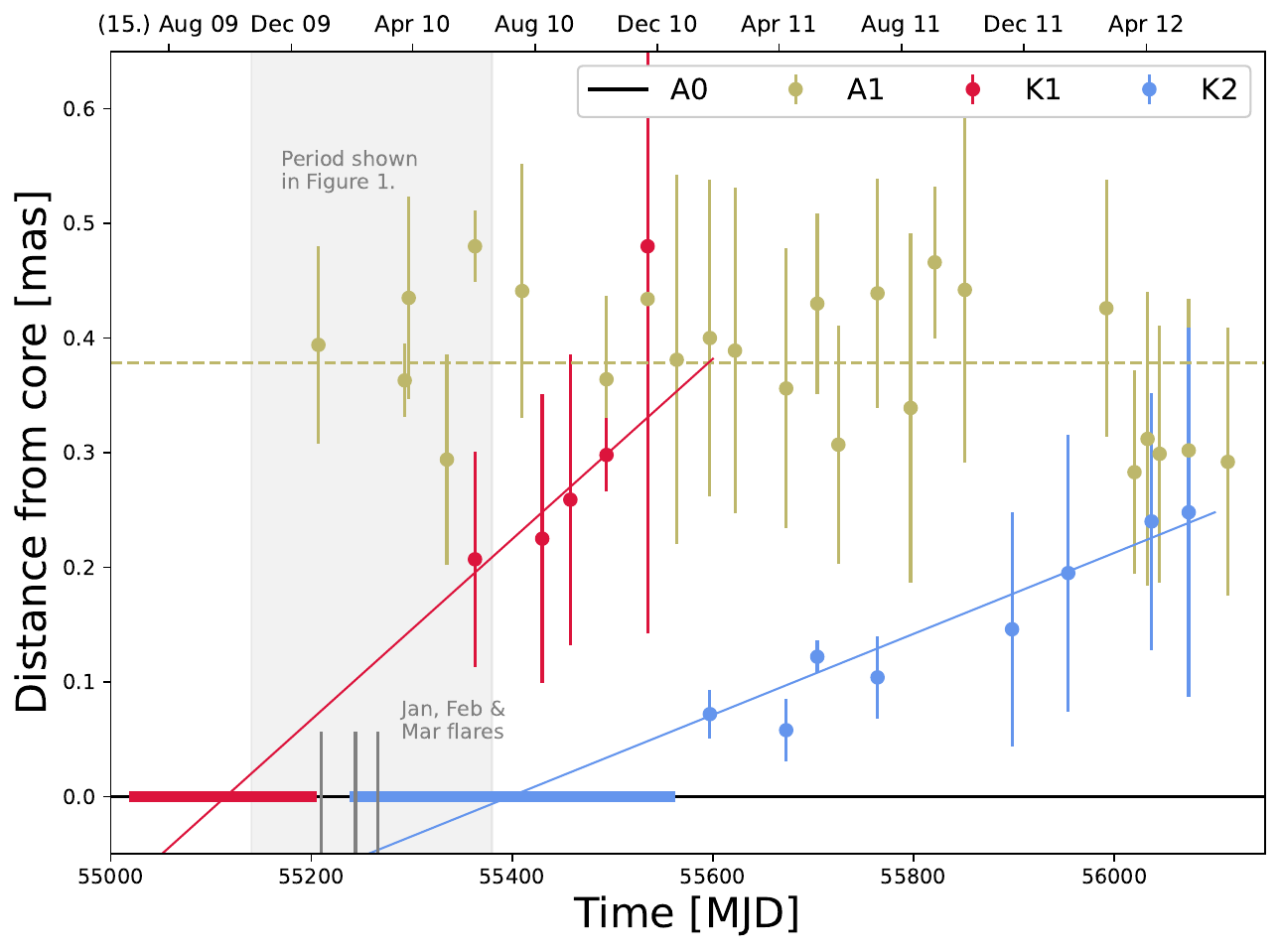}}
  \caption{Angular distance of A1 (yellow), K1 (red), and K2 (blue) from the jet core (black). At the location of Mrk\,421, the angular distance of one mas is equivalent to a physical distance of 0.638\,pc. The dotted yellow line indicates a constant fit of the quasi-stationary component. The red and blue lines show linear fits to determine the time of ejection of the components. The uncertainties on the ejection times are given by the red and blue bands. The campaign of this work is given by the light gray band with the three flares marked with gray lines.}
  \label{fig:extrapolation}
\end{figure}

Fig.~\ref{fig:VLBA} shows the total intensity VLBA images of Mrk\,421 from May 2010 to August 2011. The images are shown from May onwards when the ejection of the first feature is observed. The source structure was modeled using a number of emission components (knots) with circular Gaussian brightness distributions. Based on their parameters, we identify knots across the epochs. The brightest knot located at the southern end of the jet is the VLBI core at 43~GHz, which we place at zero distance in Fig.~\ref{fig:VLBA}. The core, designated as A0, is assumed to be a stationary physical structure of the jet. Another (quasi-)stationary feature is located at $0.38\pm0.06\,$mas from the core, labeled A1. At the distance of Mrk\,421, one can use the conversion scale 0.638 pc mas$^{-1}$\footnote{from the NASA/IPAC Extragalactic Database, \url{https://ned.ipac.caltech.edu/}} to convert the angular distance of 0.38\,mas into a physical distance of 0.24\,pc. During the observations, two new components emerged from the core, K1 and K2. The components were already reported in~\citet{2017ApJ...846...98J} (K1 and K2 are referred to as B1 and B2, respectively). As reported there, the existence of B1 (K1 in this manuscript) as a real knot is not absolutely certain. All average parameters of the components (i.e. names of the knot, number of epochs at which the knot is detected, flux density, distance from the core, position angle with respect to the core, and the size of the knot) are listed in Tab.~\ref{tab:VLBA_params}. The corresponding kinematic properties of K1 and K2 are listed in Tab.~\ref{tab:VLBA_kin} (Phi refers to the speed direction). The newly emerging component K1 moves at a speed of 0.78\,mas/yr, which corresponds to an apparent speed of 1.56$\pm$0.45\,c, making it just about a superluminal knot (note the large uncertainty). K2 moves much slower with a speed of 0.16\,mas/yr corresponding to 0.32$\pm$0.07\,c. \par
Fig.~\ref{fig:extrapolation} shows the angular distance from the core of the components A1, K1, and K2, as a function of time. Since A1 is identified as a stationary feature, the distance versus time of A1 with respect to A0 is fit with a horizontal line, while separations of two new components from the core are approximated by linear fits with the best $\chi^2$ value. This allows us to determine times of ejection (passage through the center of the core) of K1 and K2 by extrapolation of the motion of knots back to the core position (see Tab.~\ref{tab:VLBA_kin}).

K1 is ejected at MJD 55112$\pm$88 and K2 at MJD 55400$\pm$157. Both ejection times fall well within the main period of this work, shown in light gray, from MJD 55140 to 55380 within their one-sigma band. Because of the large uncertainties, however, it is impossible to establish a connection with a particular event, such as one of the three VHE flares, marked with solid gray lines. \par
Nevertheless, it is remarkable that in the period Mrk\,421 exhibited the most violent behavior with three large VHE flares, including the brightest flare ever, the ejection of two new components was seen right after. Despite VLBA components having been identified for a few TeV-emitting blazars (mostly FSRQs), a potential association of gamma-ray flares with the ejection and propagation of knots is rarely observed in HSPs~\citep{2022ApJS..260...12W}. To our knowledge, this is the only time the ejection of new components in the jet of Mrk\,421 was observed. Previous works only found stationary or already present subluminal knots~\citep[e.g.][]{2005ApJ...622..168P, 2012A&A...545A.117L, 2013EPJWC..6104010R, 2013A&A...559A..75B}. \par

\section{Variability} \label{sec:variability}

\subsection{Fractional variability} \label{sec:fvar}
\begin{figure}[h]
\centering
  \resizebox{\hsize}{!}{\includegraphics{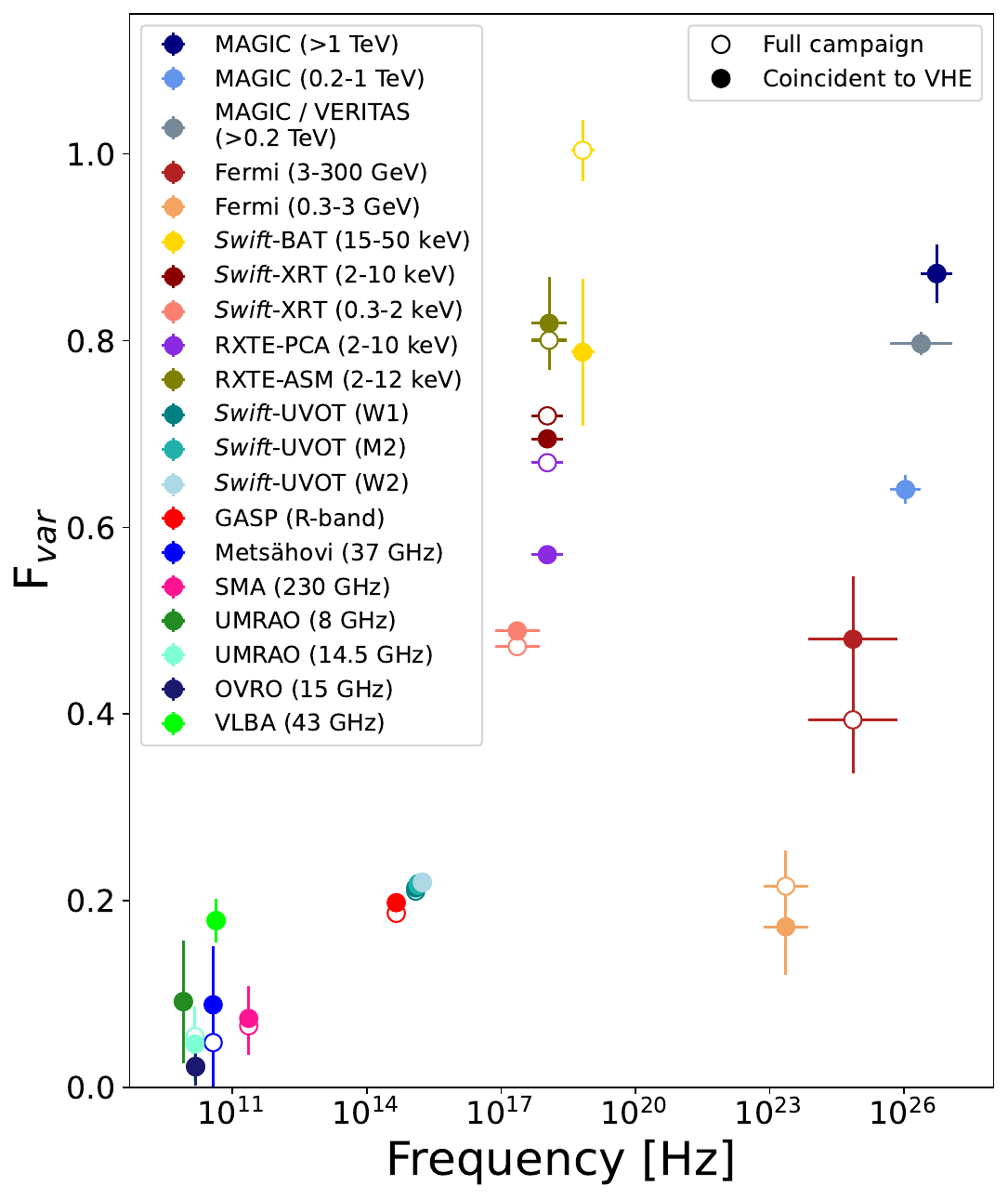}}
  \caption{Fractional variability $F_{var}$ as a function of the frequency for the light curves shown in Fig.~\ref{fig:MWL_LC}. Open markers show the results using the whole campaign for each light curve. Full markers only include quasi-simultaneous data to the VHE data (quasi-simultaneity is defined as temporal agreement with VHE data within 6\,h for X-ray and UV data, within 1 day for optical data, and within 3 days for radio and \textit{Fermi}-LAT data.)}
  \label{fig:fvar_vs_freq}
\end{figure}

In order to estimate the degree of variability in each energy band, we use the fractional variability $F_{var}$ as it is described in~\citet{2003MNRAS.345.1271V}. The estimation of the uncertainties follows the description in~\citet{2015A&A...573A..50A}, which uses the approach from~\citet{2008MNRAS.389.1427P}. Fig.~\ref{fig:fvar_vs_freq} shows the resulting fractional variability using the full light curves in open markers. The full markers show the results for quasi-simultaneous data. We define quasi-simultaneity as the temporal agreement with VHE data within 6\,h for X-ray and UV data. The choice of 6\,h is a compromise between having sufficient data and not too much loss of temporal coincidence. We use a window of 1\,day for optical data and 3\,days for radio and \textit{Fermi}-LAT data. \par

In both cases, a pronounced two-peak structure is visible, indicating the highest variability in the Xray and VHE. Similar behavior has been observed multiple times for Mrk\,421~\citep[e.g.][]{2015A&A...576A.126A, 2016ApJ...819..156B, 2021A&A...655A..89M}. As mentioned in Sec.~\ref{sec:intro}, the typical SED from a blazar such as Mrk\,421 shows a double bump structure~\citep{abdo:2011}, where the first bump is electron synchrotron emission. The falling edge of that bump, covered by the X-rays, is emitted by the most freshly accelerated and energetic electrons, which also have the shortest cooling times, meaning high variability. At energies below the X-rays, the electrons responsible for the synchrotron emission have lower energies and hence longer cooling times, resulting in a strong decrease in variability. In the context of a SSC model, the same electrons produce the second bump of the SED in the gamma rays via inverse Compton emission. A similar variability pattern is therefore also expected in the variability of the gamma-ray peak. Fig.~\ref{fig:fvar_vs_freq} shows the same degree of variability at TeV energies as in keV energies. This result already suggests the existence of a correlation between the two wavebands, which is investigated in more detail in Sec.~\ref{sec:correlation}. \par

\subsection{Power spectral density} \label{sec:PSD}
Additionally, we investigate the variability of Mrk\,421 with the use of the power spectral density (PSD). The PSD quantifies the amplitude of the variability as a function of the timescale of the variations. It is based on the discrete Fourier transform of a light curve. The typical shape of the PSD for blazars follows a simple power law $P_{\nu} \propto \nu^{-a}$ with the spectral index $a$ ranging between 1 and 2~\citep[e.g.][]{2002MNRAS.332..231U, 2008ApJ...689...79C, 2010ApJ...722..520A}. A falling power law indicates a large variability at long timescales (small frequencies), and the power of this variability decreases as one considers shorter and shorter timescales (high frequencies). \par 

Our method of estimating the PSD indices is described in Appendix~\ref{sec:fit_PSD}, and the results obtained are reported in Table~\ref{tab:PSD_Fit}. We find values for the PSD indices that are well compatible within their uncertainties with the ones derived in~\citet{2015A&A...576A.126A}. Due to the larger uncertainties of the light curve, the PSD indices for the \textit{Fermi}-LAT bands are not well constrained and show noticeably smaller values. We did not find evidence of a spectral break of the PSD index in any energy band. In addition, we used the Lomb-Scargle method~\citep{1976Ap&SS..39..447L, 1982ApJ...263..835S} to search for signs of periodicity but found no evidence in our data set. \par

\begin{center}
\begin{table}[h]
\caption{PSD index best fit $a$ for each energy band.} 
\centering
\begingroup
\setlength{\tabcolsep}{10pt} 
\renewcommand{\arraystretch}{1.5} 
\begin{tabular}{ l l}
Energy band (instrument) & Best fit $a$ \\
\hline \hline
> 1$\,$TeV (MAGIC) & 1.4$^{+0.4}_{-0.3}$ \\
0.2-1$\,$TeV (MAGIC) & 1.5$^{+0.5}_{-0.3}$ \\
> 0.2$\,$TeV (MAGIC/VERITAS) & 1.5$^{+0.5}_{-0.2}$ \\
3-300$\,$GeV (\textit{Fermi}-LAT) & 0.8$^{+1.0}_{-0.6}$ \\
0.3-3$\,$GeV (\textit{Fermi}-LAT) & 0.4$^{+0.8}_{-0.3}$ \\
15-50$\,$keV (\textit{Swift}-BAT) & 1.3$^{+0.2}_{-0.2}$ \\
2-10$\,$keV (\textit{Swift}-XRT) & 1.4$^{+0.2}_{-0.2}$ \\
0.3-2$\,$keV (\textit{Swift}-XRT) & 1.4$^{+0.2}_{-0.2}$ \\
W1 (\textit{Swift}-UVOT) & 1.6$^{+0.3}_{-0.3}$ \\
R-band (GASP-WEBT) & 1.9$^{+0.3}_{-0.2}$ \\
\hline \hline
\end{tabular}
\endgroup
\label{tab:PSD_Fit}
\end{table}
\end{center}

\section{Correlation studies} \label{sec:correlation}
\subsection{VHE gamma rays versus X-rays} \label{sec:VHE_vs_xray}

\begin{figure*}[h!]
\centering
  \resizebox{0.7\hsize}{!}{\includegraphics{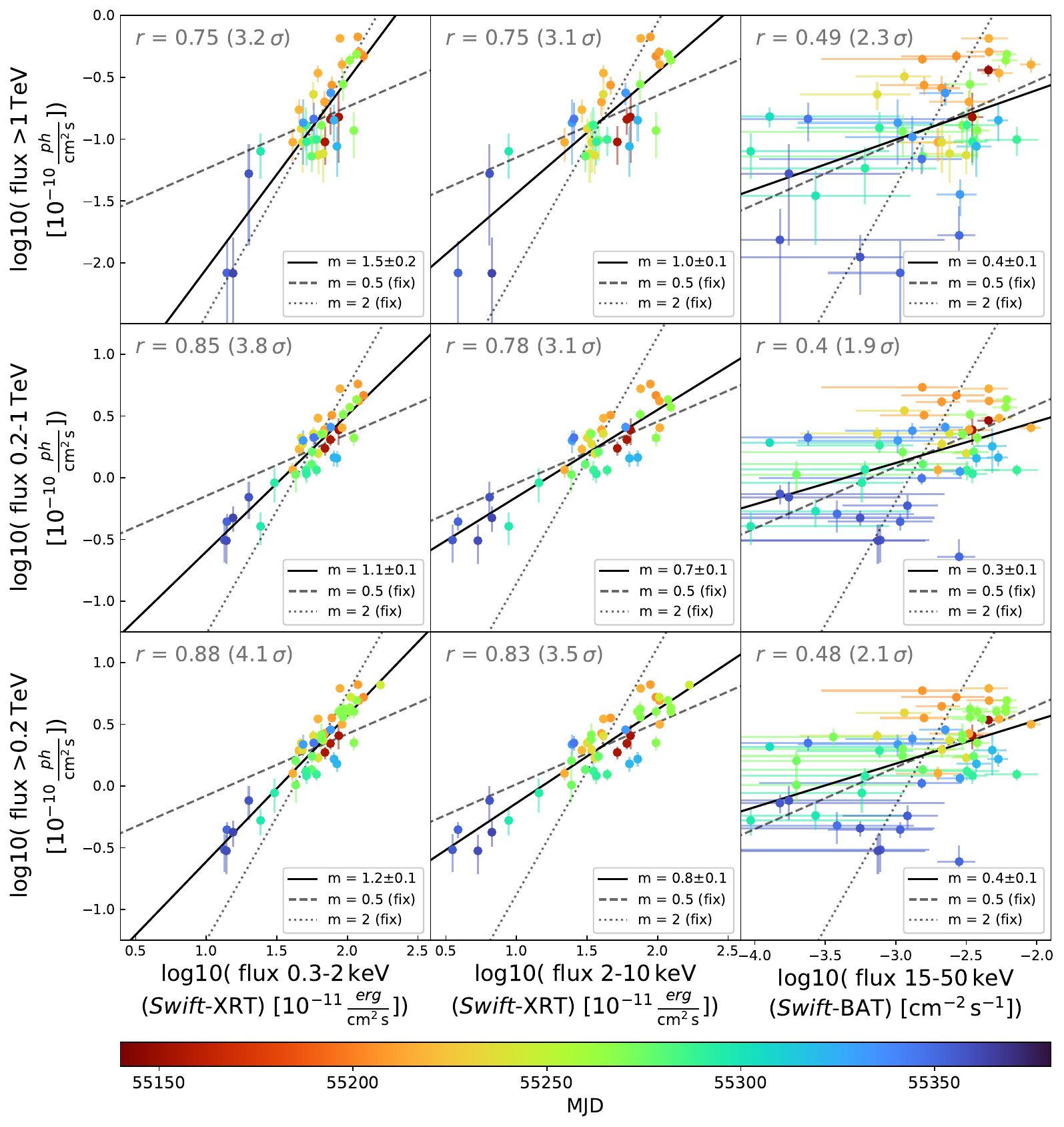}}
  \caption{VHE flux versus X-ray flux obtained by MAGIC/VERITAS and \textit{Swift}-XRT/BAT. Only pairs of observations within 6\,hours are considered. If more than one \textit{Swift} observation falls within that window, the weighted mean is computed. VHE fluxes are in the >1\,TeV band (top panels), in the 0.2-1\,TeV band (middle panels), and in the >0.2\,TeV band (bottom panels). \textit{Swift}-XRT fluxes are computed in the 0.3-2\,keV (left panels) and 2-10\,keV bands (middle panels). \textit{Swift}-BAT provides the flux in the 15-50\,keV band (right panels). The top left corner of each panel shows the Pearson coefficient of the flux pairs, with the significance of the correlations given in parentheses. The gray dashed and dotted lines depict a ﬁt with slope ﬁxed to 0.5 and 2, respectively, and the black line is the best-ﬁt line to the data, with the slope quoted in the legend at the bottom right of each panel.} 
  \label{fig:VHE_vs_xray}
\end{figure*}

The VHE gamma rays and the X-rays are the two energy bands where Mrk\,421 shows the highest flux variations (as displayed in Fig.~\ref{fig:fvar_vs_freq}), which can occur on timescales as short as hours 
(see Appendix~\ref{sec:appendix_intranight}). Because of that, the observing campaign was organized with a special focus on maximizing the simultaneity between the VHE gamma rays and X-ray measurements. Hence, the collected dataset allows us to investigate the correlations with a large number of VHE gamma rays and X-ray measurements taken within a time window of only a few hours. Previous works investigated the source behavior either during a low state of activity~\citep{2016ApJ...819..156B}, a typical state of activity~\citep{2021A&A...655A..89M} or during a flaring activity~\citep{2020ApJS..248...29A}. The large variability shown by Mrk\,421 during the MWL campaign in 2010 allows us to probe the correlation behavior across all states of emission within a single campaign lasting about half a year. Since we did not find a time delay between the two bands, we evaluate the data at zero time lag and with a flux-flux plot within a time window of only 6\,hours. Fig.~\ref{fig:VHE_vs_xray} shows the the fluxes (in decimal logarithmic scale) of all three VHE bands versus the fluxes of the two \textit{Swift}-XRT bands and the one provided by \textit{Swift}-BAT.  \par
In order to evaluate the degree of correlation, we use the Pearson coefficient using the pairs of flux data shown in Fig.~\ref{fig:VHE_vs_xray}, but without applying the decimal logarithm. We quantify the significance of the correlation using dedicated Monte-Carlo simulations. The methods are described in Appendix~\ref{sec:significances_scatter}. We find strong correlations between all VHE and \textit{Swift}-XRT energy bands. In all six panels, we find Pearson coefficients equal or above 0.75 with significances ranging from slightly above 3 up to 4\,$\sigma$. The correlation between the VHE and the \textit{Swift}-BAT band is much weaker, reaching coefficients around 0.4-0.5 and significances around 2\,$\sigma$. The significance of the correlation may be reduced by the large flux uncertainties of the BAT measurements (in comparison to the small uncertainties in the flux measurements from XRT).\par

We investigate the correlation slopes by fitting lines to the decimal logarithm of the data shown in the scatter plots. Since we did not find a significant change in the correlation slope over time by fitting individual $\sim$1 month periods, we show a single line fit for the whole campaign. It is noteworthy that in the subplots between the >0.2\,TeV flux with both X-ray bands, the observations of the highest and lowest X-ray flux lie well on the linear fit. This indicates a linear trend reaching over a full order of magnitude of emission and on the timescale of multiple months. The steepest slope ($1.5\pm0.2$) is obtained for the highest VHE gamma-ray band versus the lowest X-ray band, while the flattest slope ($0.3\pm0.1$) is obtained for the lowest VHE gamma-ray band versus highest X-ray band. Overall, the slope is increasing with rising VHE energy, showing a greater scaling of the >1\,TeV flux with rising X-ray fluxes. Additionally, the slope decreases with rising X-ray energy. The same trends have been found in~\citet{2021A&A...655A..89M} as well, but the absolute slope values were found to be greater, reaching a cubic relation in some cases. \par

\begin{figure}[h!]
\centering
  \resizebox{0.8\hsize}{!}{\includegraphics{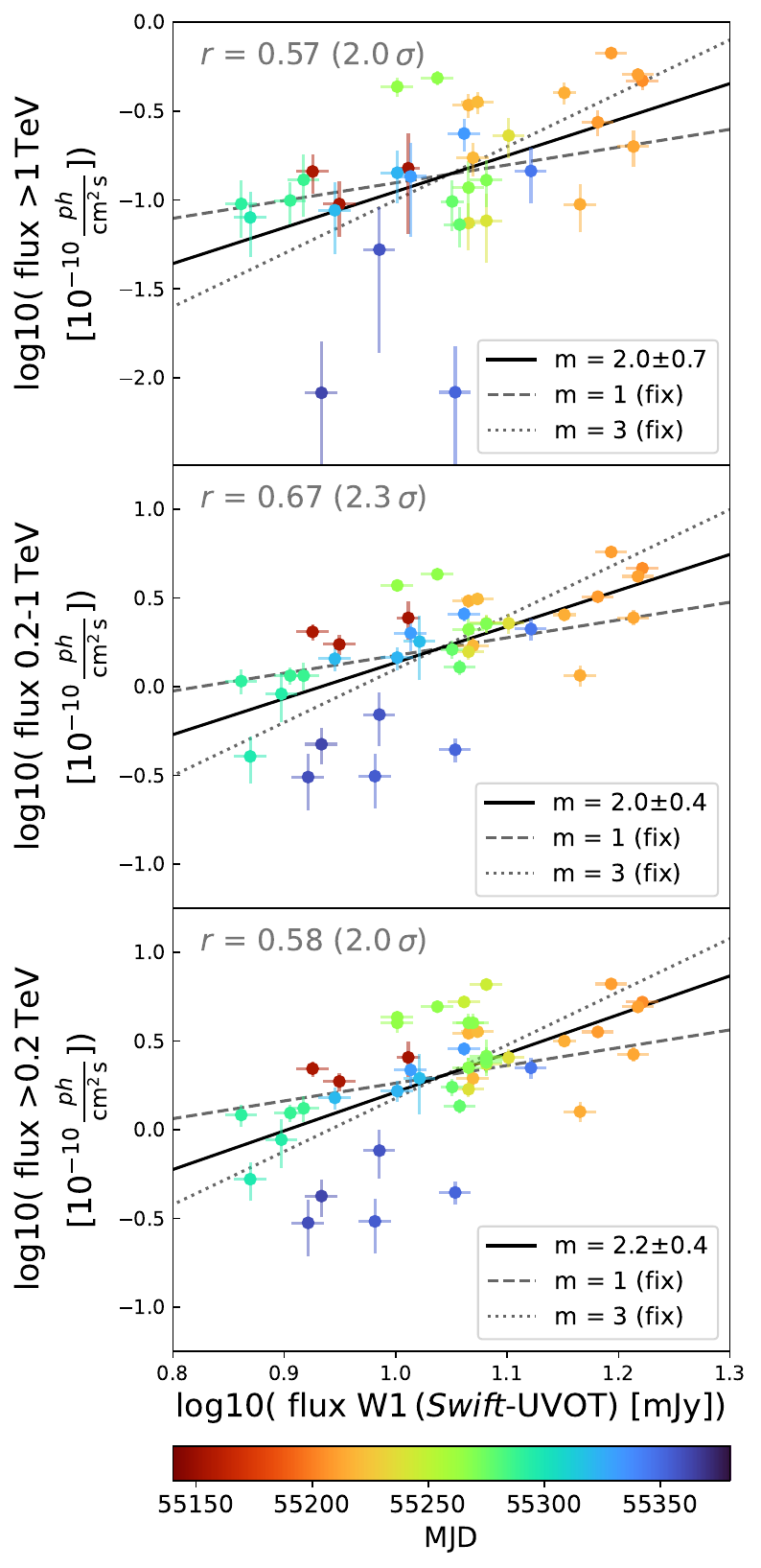}}
  \caption{VHE flux versus UV flux obtained by MAGIC/VERITAS and \textit{Swift}-UVOT. Only pairs of observations within 6\,hours are considered. If more than one \textit{Swift} observation falls within that window, the weighted mean is computed. VHE fluxes are in the >1\,TeV band (top panels), in the 0.2-1\,TeV band (middle panels) and in the >0.2\,TeV band (bottom panels). The \textit{Swift}-UVOT fluxes are taken with the W1 filter. The top left corner of each panel shows the Pearson coefficient of the flux pairs, with the significance of the correlations given in parentheses. The gray dashed and dotted lines depict a ﬁt with slope ﬁxed to 0.5 and 2, respectively, and the black line is the best-ﬁt line to the data, with the slope quoted in the legend at the bottom right of each panel.}
  \label{fig:VHE_vs_UV}
\end{figure}

\subsection{VHE gamma rays versus UV}
The \textit{Swift}-UVOT instrument has the same data coverage in the UV as \textit{Swift}-XRT in X-rays. This allows us to follow the same approach as in the previous section to investigate a possible correlation between the VHE and UV band pairs. 
Fig.~\ref{fig:VHE_vs_UV} shows the decimal logarithm of the fluxes of all three VHE bands versus the flux of the W1 filter (all three UV filters provide almost identical results, and we have selected W1 as a representative for all correlations). The values for the Pearson coefficient range from around 0.45 to 0.65 for all three panels, indicating a much weaker correlation. However, the corresponding significances of these correlations are all around 2$\,\sigma$, and hence, this result should be considered as a hint of correlation.
If the correlation were real, this would be the first time, to our knowledge, that a positive correlation between VHE gamma rays and the UV band is found. \par
For the linear fits, we obtain compatible values of around m=2 for all three panels. Due to the higher dispersion of the data, the uncertainties of the fit are rather large. The higher slope indicates a steeper correlation of the VHE with UV compared to X-rays. Since the degree of variability is much higher in the VHE than the UV (see Fig.~\ref{fig:fvar_vs_freq}), the flux varies substantially more, and a steeper slope is expected in the case of a correlation. \par

\subsection{VHE versus HE gamma rays} \label{sec:correlation_VHE_vs_HE}
Next, we investigate the correlation between the two energy bands in gamma rays using the combined light curve above 0.2\,TeV from MAGIC and VERITAS. We correlate the bands using the DCF without rebinning the LCs. We use a 3-day binned time lag in the range of -40 to +40 days. The significance of the resulting DCF is estimated with simulations similar to the approach in~\citet{2021A&A...655A..89M}: We first simulate a set of 10000 uncorrelated light curves for a pair of energy bands as before. We then compute the DCF of the 10000 simulated light curve pairs. The $1\,\sigma$, $2\,\sigma$ and $3\,\sigma$ confidence bands are derived from the distribution of the simulated DCF values in each time lag bin.\par 

The result for the DCFs between the >0.2\,TeV band with the 3-300\,GeV band is shown in Fig.~\ref{fig:VHE_vs_fermi_HE}. The figure shows the DCF obtained from the data in dark blue. The DCF peaks at a time lag of $2\pm2$\,days, where it crosses the 3\,$\sigma$ line. To estimate the uncertainty of the time lag, we follow the method outlined in~\citet{t_lag_uncertainties}. \par
We note that the marginally significant (a little over 3$\,\sigma$) peak at zero time lag is primarily driven by the big flare in February 2010 because the highest flux points in the VHE and HE light curves occur during this flare. If the highest flux in the HE light curve (that relates to a 3-day time interval from MJD 55242.0 to MJD 55245.0 or 2010-02-15 00:00 to 2010-02-18 00:00), together with the corresponding VHE fluxes in this 3-day time interval, are removed from the DCF study, the correlation at zero time lag vanishes, as shown in Fig.~\ref{fig:VHE_vs_fermi_HE_noflare}. A similar behavior is seen in the correlation between the >0.2\,TeV band and the 0.3-3\,GeV band, as it is also shown in Figs.~\ref{fig:VHE_vs_Fermi_LE} and~\ref{fig:VHE_vs_Fermi_LE_noflare}. But this time, the significance of the peak at zero time lag is just over the 2\,$\sigma$ line, even including the 2010 February flare. \par

\subsection{HE gamma rays versus X-rays}
\label{sec:HEvsX}
For the correlation between HE gamma rays from \textit{Fermi}-LAT and X-rays from \textit{Swift}-XRT, we again use the DCF. We correlate each of the two HE gamma-ray bands with each of the X-ray bands from \textit{Swift}-XRT. Here, we only display the strongest correlation as an example, which is found between the 3-300\,GeV and the 0.3-2\,keV band, shown in Fig.~\ref{fig:Fermi_HE_vs_Xray_LE}. A correlation peak lying above 3$\,\sigma$ is found at a time lag of $-1\pm2\,$days. Similar to the VHE versus HE gamma-ray case, the sharp peak at zero time lag is likely dominated by the outstanding flaring activity on February 17. Fig.~\ref{fig:Fermi_HE_vs_Xray_LE_noflare} shows the same correlation but with the flare removed from both the light curves taken by \textit{Fermi}-LAT, as well as the simultaneous flux in the X-ray light curve. The high peak at exactly zero time lag vanishes, but a broad peak remains. \par
The other three combinations of energy bands (0.3-3\,GeV versus 0.2-3\,keV, 3-300\,GeV versus 2-10\,keV, 0.3-3\,GeV versus 2-10\,keV) are also shown in Fig.~\ref{fig:Fermi_Xray}. In all combinations, one can see a marginally significant peak at a time lag of about zero that disappears when removing the 3-day LAT flux bin (and corresponding X-ray fluxes) from the outstanding flaring activity around 2010 February 17. The period in early 2010, including the flare, marks the first time such a correlation was observed, as previously reported in~\citet{2020ApJ...890...97A}. 

\subsection{HE gamma rays versus UV}
Fig.~\ref{fig:Fermi_HE_vs_UV} shows the DCF between the 3-300\,GeV and the UV bands. It exhibits a broad peak crossing the 3$\,\sigma$ line, and it reaches its highest value at a time lag of $-7\pm6$\,days. Between the 0.3-3\,GeV band and the UV in Fig.~\ref{fig:Fermi_LE_vs_UV}, the highest value at $-1\pm6$\,days also reaches a significance level above 3$\,\sigma$. The positive correlation between both HE gamma-ray regions and the UV band is a novel behavior for Mrk\,421 with no previous reports to our knowledge. \citet{2021MNRAS.504.1427A} found a positive correlation compatible with a time lag of zero between the 0.3-300\,GeV band, and the R-band, which frequency is close to the UV. The correlation was found by investigating the time period 2007-2016 with a 15-day binning resulting in a lower temporal resolution than the one used here. \par
Contrary to the previous cases, the correlation between HE gamma-rays and the UV increases if the flare in February is excluded. Figs.~\ref{fig:Fermi_HE_vs_UV_noflare} and ~\ref{fig:Fermi_LE_vs_UV_noflare} show the results of correlating the HE gamma-ray light curves with the UV light curve without the flare. In the case of 3-300\,GeV, the peak is clearly enlarged, going well above 3\,$\sigma$, indicating a stronger correlation than before. For 0.3-3\,GeV, the improvement is only marginal. The peaks remain at the same time lag as before. \par 

\subsection{HE gamma rays versus R-band}
We additionally test whether the correlation between the HE gamma rays and the optical as previously reported in~\citet{2021MNRAS.504.1427A} is also present in our data set. Fig.~\ref{fig:Fermi_vs_optical} shows the DCF for the R-band and the two HE bands, which does not lead to any significantly correlated behavior, apart from a marginally significant ($\sim 3\sigma$) feature at about -35 days. \par
Removing the 3-day interval that contains the large VHE and X-ray flare in February 2010 does not change substantially the shape and the significance of the DCF results for these two bands. Only a marginal increase to 2.5$\,\sigma$ is seen for the 3-300\,GeV versus optical (see Fig.~\ref{fig:Fermi_HE_vs_optical_noflare}).  \par

\subsection{Other energy bands}
We also investigated the correlation behavior between all other wavebands not mentioned in the previous sections. We found no correlation (positive or negative) between any bands in the \mbox{X-ray} with the UV or the optical. Previous works have found anti-correlated behavior at times~\citep[e.g.][]{2021A&A...655A..89M, mrk421_IXPE_MAGIC}, which was not present in our data set. Naturally, there is a strong correlation between the UV and optical, as the energy is close to each other. Since we established a hint of a correlation between VHE and UV, we also checked if a correlation is present between the VHE and the optical. Applying the same method, we find no correlations at all between VHE and optical. We also found no correlation of any sort between radio and other energy bands. Due to the limited duration of the campaign (about seven months), the overall number of radio observations is also relatively small, in comparison to studies that used multi-year datasets \citep[see e.g.][]{2017MNRAS.472.3789C,2021MNRAS.504.1427A}. In addition, the synchrotron self-absorption of radio radiation might reduce any detectable correlation. \par

\section{Discussion} \label{sec:discussion}
\subsection{First detection of ejection of radio knots contemporaneous to a flare of Mrk~421}
\label{sec:disscussion-VLBA}
VLBA observations reveal the ejection of two features in the jet that are close in time to the observed VHE gamma-ray flares (see Sect.~\ref{sec:VLBA}). While the association of the ejection and propagation of bright knots in the jet with gamma-ray flares has been observed repeatedly in blazars, the vast majority of these detections were reported in FSRQs (like PKS~1510-089) or LSP and ISP BL Lacertae \citep[see][]{2001ApJ...556..738J, 2008Natur.452..966M, 2018A&A...619A..45M}. The association has rarely been observed in HSPs, and this is the first time for Mrk~421~\citep[see e.g. the sample in][]{2022ApJS..260...12W}, hence making the dataset presented in this paper unique on its own. However, the large uncertainties of the ejection time prevent us from significantly correlating the appearance of the radio features with the specific VHE gamma-ray flares during this campaign. We also note that previous works on Mrk\,421 reported a possible connection between GeV gamma-ray and radio flares detected with single-dish telescopes ~\citep{2015MNRAS.448.3121H}. However, this large GeV flaring activity, which was measured with \textit{Fermi}-LAT in the year 2012, and the exceptionally large radio flare, which was (mostly) measured with OVRO also in 2012, was not associated with the ejection of features in the VLBA images~\citep{2013EPJWC..6104010R}. Hence, they can have a different nature, and are not comparable to the ejection of features in the VLBA images during the 2010 campaign that is reported in this manuscript. \par

Assuming the flare in February is indeed correlated with the ejection of the radio features, the VHE intranight variability implies a small, compact zone responsible for the emission based on causality arguments. As discussed in \citet{2020ApJ...890...97A}, following~\citet{1995MNRAS.273..583D}, the intranight variability down to 22\,min at VHE in February 2010 allows us to derive a lower limit on the Doppler factor of $\delta_{\text{min}}\gtrsim33$ based on opacity arguments.\par

\citet{2017ApJ...846...98J} estimate the average Doppler factor derived from VLBA observations to be around 24, hence below the lower limit mentioned above using the VHE observations. This regularly observed discrepancy between the observed jet kinematics in the radio and the Doppler factors derived from the fast variability at high energies is referred to as the \textit{Doppler crisis}. This could potentially be resolved by the presence of structured jets, in which different regions show different Lorentz factors, such as a fast spine responsible for the gamma-ray emission and a slower sheath layer responsible for the radio emission~\citep{2005A&A...432..401G}. Recent works also proposed a scenario in which the discrepancy is reconciled by assuming, on the one hand, a large viewing angle relative to the jet explaining the slow-moving knot in radio, while on the other hand, the gamma-ray flare originates from a magnetic reconnection event within a misaligned layer that effectively generates large Doppler factors~\citep{2023A&A...678A.140J}. Finally, jet deceleration occurring between the time of the gamma-ray flare and the detection of the radio feature may partly solve this crisis \citep{2003ApJ...594L..27G}. \par

The FWHM of the radio knots reported in Sect.~\ref{sec:VLBA} is in the order of 0.15\,mas (both for K1 and K2). At the distance of Mrk~421, this corresponds to a linear size of roughly 0.1\,pc$\approx 3\times10^{17}$\,cm (0.638\,pc/mas). Assuming a Doppler factor of 33 and a variability timescale of 22 minutes, the upper limit to the size of the emitting region during the VHE flare is in the order of $10^{15}$\,cm~\citep{2020ApJ...890...97A}. If one again postulates that the radio knot corresponds to the same region that produced the VHE flare in February 2010, this implies an expansion velocity in the comoving frame of the emitting region in the order of $\beta_{exp}=10^{-2}c$ (using a time difference between the gamma-ray flare and the radio detection of roughly 300\,days, as one can infer from Fig.~\ref{fig:extrapolation}).  It is interesting to note that such expansion velocity is well in line with the estimates of \citet{2022A&A...658A.173T}, which are obtained when trying to explain the delayed gamma-ray and radio response in Mrk~421 by an adiabatic expansion of the blob. The adiabatic expansion of the blob is expected to induce a decrease in the electron density on timescales of $T_{ad}(t) = \frac{R(t)}{3\,\beta_{exp}\,c}$ \citep{1975ApJ...196..689G}. At the time of the VHE flares $R \sim 10^{15}$\,cm, hence $T_{ad} \approx 3\times 10^2$\,hrs, corresponding to $\approx 10$\,hrs in the observers frame, being much longer than the minimum variability timescale noted at VHE (22\,min). Hence, the VHE flux variability at the shortest timescales is likely dominantly driven by acceleration and cooling mechanisms and negligibly affected by the expansion of the blob.

\subsection{Multiband correlations as a probe of SSC in the jet}
\label{sec:discussion-VHE-X-ray}
We find a close correlation between the emission in the VHE gamma rays at zero time delay and the X-rays over an order of magnitude in flux. We also find a comparable level of variability in both energy bands, similar to the results in ~\citet{2021A&A...655A..89M} and~\citet{2021MNRAS.504.1427A}. The tight correlation and similar variability behavior suggest a cospatial origin, which is in good agreement with the SSC scenario~\citep[e.g.][]{1992ApJ...397L...5M}. The strongest correlation is found between the 0.3-2\,keV and >200\,GeV bands reaching a level of 4\,$\sigma$. Assuming standard values for the jet parameters ($B=0.1$\,G and $\delta=35$~\citep[see e.g.][]{2015A&A...578A..22A}), one derives that electrons with Lorentz factor of around $10^5$ (source reference frame) are required to emit $\sim$\,keV photons (in observer's frame). Using Eq.~14 of \citet{1998ApJ...509..608T}, which takes into account the Klein-Nishina effects, one expects that electrons with such energy would emit $\sim$0.5\,TeV photons via IC scattering of $\sim$\,keV photons. This is well in agreement with the correlation trends. \par

We also report marginally significant correlations between the UV and HE gamma rays for the first time in Mrk~421, with both bands showing a similar degree of variability. The pattern is again consistent with the SSC model, but in this case caused by electrons of lower energies populating a larger emission zone. Following an analogous approach as in the previous case, we derive estimates of the necessary electron Lorentz factors. Synchrotron photons in the UV W1 band are produced by electrons with a Lorentz factor of $\sim$$10^4$. These electrons produce IC emission at a few tens of GeV, falling well within the 3-300\,GeV band, for which the highest correlation is observed. \par

The correlation strength decreases to a non-significant level when going from the UV to the optical band. Even though the UV and R-band are close together in frequency, these results imply that the underlying particle population responsible for the HE gamma rays are dominantly radiating in the UV rather than in the R-band (assuming a leptonic model). ~\citet{2017MNRAS.472.3789C} and~\citet{2021MNRAS.504.1427A} reported significant positive correlations between the HE gamma rays and the R-band. These two studies considered much longer periods ($\sim$8 years) where, in relative terms, Mrk\,421 showed overall lower activity. The slightly different correlation behavior might be explained by the different overall magnitude and timescales of the flux variability considered, as well as the underlying mechanisms driving the emission variability (such as the evolution of the emitting region environment, change in the acceleration and cooling efficiencies etc...) \par

\subsection{Evidence of multiple emission zones}
\label{sec:discussion-multizone}
We find no correlation between the X-ray and the UV. The most simple explanation of this result is with the existence of two separate particle populations, of which a compact zone with higher particle energies is responsible for the keV (and TeV) emission, and a larger and more extended zone dominates the emission in the eV (and possibly GeV) band.
If the compact zone is embedded into the larger zone, this two-zone scenario would also be consistent with the marginal significant ($\sim$$2.0-2.3\,\sigma$) correlation between VHE gamma rays and the UV. Should the UV emission increase due to a change in the underlying particle population, synchrotron photons from the extended zone would enter the compact zone and provide additional seed photons, which would then be IC scattered and hence produce an increase in the observed VHE gamma-ray flux.  \par

The preference for a multiple-zone scenario was also noted in earlier works that included X-ray data from this campaign~\citep[see e.g.][]{2018ApJ...858...68K}. It is also further motivated by new findings of the Imaging X-ray Polarimetry Explorer (\textit{IXPE}), which suggest an energy dependency of the polarization degree from radio to the X-ray. The energy dependency points towards an energy-stratified jet creating emission zones of different extent~\citep{IXPE_Mrk421,mrk421_ixpe_rotation,mrk421_IXPE_MAGIC}. The proposed scenario of two separate emission zones is a simplified implementation of an energy-stratified jet. Similar polarization results have also been found for the HSP Mrk\,501~\citep{2022Natur.611..677L}. The outlined two-zone leptonic model with interaction between the zones has been successfully used to fit the SEDs during the \textit{IXPE} observations~\citep{mrk501_IXPE_MAGIC}. \par 

\subsection{Peculiarities of the exceptional flare in February 2010}
\label{sec:discussion-flareFeb2010}
We find a positive correlation between the VHE and HE gamma rays over the whole campaign. A similar correlation was first reported for a dataset taken in \mbox{2015-2016,} during which Mrk\,421 was in a historically low-activity state~\citep{2021MNRAS.504.1427A}. However, the correlation in our data set is non-significant when removing the 3-day interval related to the large February 2010 flare, and hence these two bands were directly connected only during this exceptional flaring event. In other words, the VHE versus HE correlation is not representative of the behavior of Mrk\,421 during the \mbox{8-month} long 2010 campaign (which covers different activity levels).\par 
The same phenomenon is observed when considering the HE gamma rays and the X-rays, for which the correlation also becomes non-significant when the 3-day time interval of the February 2010 flare is removed. Including the flare, we find the strongest correlation for the highest-energy band of \textit{Fermi}-LAT (3-300\,GeV) and the lowest-energy band of \textit{Swift}-XRT (0.3-2\,keV). While our study shows that the X-ray and VHE radiation share a common emission region and the UV and HE bands may originate from a different region, the flare could be driven by a single compact zone whose activity is so high that it not only dominates in the VHE radiation, but also contributes significantly to the emission of HE gamma rays on these days. The X-ray and HE gamma-ray bands might then be close to the peak of the SED bumps caused by synchrotron and IC emission of the same particles. Matching this scenario, the 0.3-3\,GeV band, sitting at the rising edge of the second bump, and the 2-10\,keV band, sitting at the falling edge of the synchrotron bump, show the weakest correlation.\par 

\begin{acknowledgements}
Author contribution: J. Abhir: MAGIC analysis cross-check; A. Arbet-Engels: variability and correlation analysis, discussion and interpretation, paper drafting; D. Paneque: coordination of MWL observations and coordination of the MWL data reduction, \textit{RXTE}-PCA analysis, discussion and interpretation, paper drafting; F. Schmuckermaier: project management, MAGIC and Fermi-LAT data analysis, variability and correlation analysis, discussion and interpretation, paper drafting; The rest of the authors have contributed in one or several of the following ways: design, construction, maintenance and operation of the instrument(s) used to acquire the data; preparation and/or evaluation of the observation proposals; data acquisition, processing, calibration and/or reduction; production of analysis tools and/or related Monte Carlo simulations; overall discussions about the contents of the draft, as well as related refinements in the descriptions. \\

The MAGIC collaboration would like to thank the Instituto de Astrof\'{\i}sica de Canarias for the excellent working conditions at the Observatorio del Roque de los Muchachos in La Palma. The financial support of the German BMBF, MPG and HGF; the Italian INFN and INAF; the Swiss National Fund SNF; the grants PID2019-104114RB-C31, PID2019-104114RB-C32, PID2019-104114RB-C33, PID2019-105510GB-C31, PID2019-107847RB-C41, PID2019-107847RB-C42, PID2019-107847RB-C44, PID2019-107988GB-C22, PID2022-136828NB-C41, PID2022-137810NB-C22, PID2022-138172NB-C41, PID2022-138172NB-C42, PID2022-138172NB-C43, PID2022-139117NB-C41, PID2022-139117NB-C42, PID2022-139117NB-C43, PID2022-139117NB-C44 funded by the Spanish MCIN/AEI/ 10.13039/501100011033 and “ERDF A way of making Europe”; the Indian Department of Atomic Energy; the Japanese ICRR, the University of Tokyo, JSPS, and MEXT; the Bulgarian Ministry of Education and Science, National RI Roadmap Project DO1-400/18.12.2020 and the Academy of Finland grant nr. 320045 is gratefully acknowledged. This work was also been supported by Centros de Excelencia ``Severo Ochoa'' y Unidades ``Mar\'{\i}a de Maeztu'' program of the Spanish MCIN/AEI/ 10.13039/501100011033 (CEX2019-000920-S, CEX2019-000918-M, CEX2021-001131-S) and by the CERCA institution and grants 2021SGR00426 and 2021SGR00773 of the Generalitat de Catalunya; by the Croatian Science Foundation (HrZZ) Project IP-2022-10-4595 and the University of Rijeka Project uniri-prirod-18-48; by the Deutsche Forschungsgemeinschaft (SFB1491) and by the Lamarr-Institute for Machine Learning and Artificial Intelligence; by the Polish Ministry Of Education and Science grant No. 2021/WK/08; and by the Brazilian MCTIC, CNPq and FAPERJ. \\

The \textit{Fermi} LAT Collaboration acknowledges generous ongoing support
from a number of agencies and institutes that have supported both the
development and the operation of the LAT as well as scientific data analysis.
These include the National Aeronautics and Space Administration and the
Department of Energy in the United States, the Commissariat \`a l'Energie Atomique
and the Centre National de la Recherche Scientifique / Institut National de Physique
Nucl\'eaire et de Physique des Particules in France, the Agenzia Spaziale Italiana
and the Istituto Nazionale di Fisica Nucleare in Italy, the Ministry of Education,
Culture, Sports, Science and Technology (MEXT), High Energy Accelerator Research
Organization (KEK) and Japan Aerospace Exploration Agency (JAXA) in Japan, and
the K.~A.~Wallenberg Foundation, the Swedish Research Council and the
Swedish National Space Board in Sweden.

Additional support for science analysis during the operations phase is gratefully 
acknowledged from the Istituto Nazionale di Astrofisica in Italy and the Centre 
National d'\'Etudes Spatiales in France. This work performed in part under DOE 
Contract DE-AC02-76SF00515.

A.A.E. and D.P. acknowledge support from the Deutsche Forschungsgemeinschaft (DFG; German Research Foundation) under Germany’s Excellence Strategy \mbox{EXC-2094–390783311}. \\

The research at Boston University was supported in part by several NASA Fermi Guest Investigator grants, the latest of which are 80NSSC23K1507 and 80NSSC23K1508. This study was based in part on observations conducted using the 1.8m Perkins Telescope Observatory (PTO) in Arizona, which is owned and operated by Boston University. The VLBA is an instrument of the National Radio Astronomy Observatory. The National Radio Astronomy Observatory is a facility of the National Science Foundation operated by Associated Universities, Inc. 
Research at UMRAO was supported by a series of grants from the NSF (most recently  AST-0607523) and from NASA (including Fermi G.I. awards NNX09AU16G, NNX10AP16G, NNX11AO13G, and NNX13AP18G).  Funds for the operation of UMRAO were provided by the University of Michigan.
The Submillimeter Array is a joint project between the Smithsonian Astrophysical Observatory and the Academia Sinica Institute of Astronomy and Astrophysics and is funded by the Smithsonian Institution and the Academia Sinica.
We recognize that Maunakea is a culturally important site for the indigenous Hawaiian people; we are privileged to study the cosmos from its summit.

\end{acknowledgements}

\bibliographystyle{aa}

\bibliography{main}

\begin{appendix}
\section{Observations and data processing} \label{sec:data_analysis}
This section contains a detailed description of all used instruments and their corresponding data analysis. Since the same campaign is exploited in parts in~\citet{2015A&A...578A..22A}, some instrument sections closely follow the descriptions given in there. \par

\subsection{VHE gamma rays}
The MAGIC telescopes are two Imaging Atmospheric Cherenkov Telescopes (IACTs), MAGIC I and MAGIC II, each with a diameter of 17$\,$ m, located at Observatorio del Roque de los Muchachos (ORM, 28.762$^\circ$N 17.890$^\circ$W, 2200 m above sea level) on the Canary Island of La Palma. With the start of stereoscopic observations in 2009 and substantial hardware upgrades completed in 2012, MAGIC is capable of detecting gamma rays with energies from about 30$\,$GeV up to $\gtrsim$100$\,$TeV~\citep{aleksic:2016, MAGICCrab100TeV}.\par 
This work covers the time period from November 11, 2009, (MJD 55149) until June 16, 2010, (MJD 55363). In total, the MAGIC telescopes observed Mrk$\,$421 for 62.4$\,$h in the zenith angle range between 5$^\circ$ and 50$^\circ$. The data are analyzed using the MAGIC Analysis and Reconstruction Software, MARS~\citep{MARS,aleksic:2016}. The final data are selected based on quality criteria to exclude periods with unfavorable weather conditions or too bright moon. The data were taken under low moonlight conditions in order to limit contamination from night sky background light \citep{2017APh....94...29A}. After applying quality cuts, 50.1$\,$h of data remained. \par 
This work focuses on the MWL variability and correlations, and hence, we use light curves that report the VHE gamma-ray emission of Mrk\,421.  Due to the high gamma-ray brightness of Mrk\,421, we can construct two separate light curves in the VHE band covering two energy ranges: 0.2-1$\,$TeV and >1$\,$TeV. The analysis was performed for the whole campaign including the already published periods in \cite{2015A&A...578A..22A} and \cite{2020ApJ...890...97A}. This ensures that the low-level analysis remains consistent throughout and the results are well compatible. In addition to the data from MAGIC, we also used VERITAS light curves with VHE gamma-ray fluxes above 0.2 TeV published in \cite{2015A&A...578A..22A} and \cite{2020ApJ...890...97A}.  In order to create a joint light curve, a third light curve was constructed from the MAGIC data for this energy range. \par

\subsection{HE gamma rays}
The \textit{Fermi Gamma-ray Space Telescope} satellite carries the LAT detector on board. It is a pair-conversion telescope surveying the gamma-ray sky in the 20\,MeV to $>300$\,GeV energy range~\citep{2009ApJ...697.1071A,2012ApJS..203....4A}. For the construction of light curves, we perform a binned-likelihood analysis using tools from the \texttt{FERMITOOLS} software\footnote{\url{https://fermi.gsfc.nasa.gov/ssc/data/analysis/}} v2.2.0. We use the instrument response function \texttt{P8R3\_SOURCE\_V3} and the diffuse background models\footnote{\url{http://fermi.gsfc.nasa.gov/ssc/data/access/lat/\\BackgroundModels.html}} \texttt{gll\_iem\_v07} and \texttt{iso\_P8R3\_SOURCE\_V3\_v1}.\par 
We create two light curves in the range from 0.3\,GeV to 3\,GeV and 3\,GeV to 300\,GeV by selecting \texttt{Source} class events in a circular region of interest (ROI) with a radius of $20^\circ$ around Mrk\,421 in the respective energy band. All events with a zenith angle above $90^\circ$ are discarded to reduce the contribution from limb gamma rays. The analysis threshold energy of 0.3$\,$GeV was chosen over the more usual 0.1$\,$GeV in order to make use of the improved angular resolution of \textit{Fermi}-LAT at higher energies. A higher energy threshold additionally reduces background contamination, which leads to an overall improvement of the signal-to-noise ratio for hard sources (photon index~$<2$) such as Mrk$\,$421. For the source model, we include all sources from the fourth Fermi-LAT source catalog Data Release 3 \citep[4FGL-DR3;][]{2020ApJS..247...33A, 2022ApJS..263...24A} that are found within the ROI plus an additional $5^\circ$. We fit the obtained model to our data covering the time period from November 2, 2009, (MJD 55141) to July 2, 2010, (MJD 55379). The initial fit results are used to remove weak components from the model (counts~$<1$ or TS~$<3$\footnote{The test statistic TS is defined as $-2 \text{log}(\mathcal{L}_{\text{max},0}/\mathcal{L}_{\text{max},1})$, where $\mathcal{L}_{\text{max},0}$ is the maximum likelihood value for a model without an additional source, the null hypothesis, and $\mathcal{L}_{\text{max},1}$ is the maximum likelihood value for a model with the additional source at a specified location. }). After the first optimization, each time bin is fit with the model. During the fitting procedure, the normalization of bright sources (TS~$>10$), sources close to the ROI center ($<3^\circ$), the diffuse background, and the spectral parameters of Mrk$\,$421 itself are allowed to vary. We produce the light curves with a 3-day binning. In all time bins, the source is detected with TS~$>25$ (i.e., $>5\sigma$). \par

\subsection{X-rays}
We scheduled observations with the X-Ray Telescope~\citep[XRT;][]{2005SSRv..120..165B} on board the \textit{Neil Gehrels Swift Observatory (Swift)} throughout the full campaign to achieve the best possible coincidence with VHE observations. All \textit{Swift}-XRT observations were carried out both in Windowed Timing (WT) and Photon Counting (PC) readout modes. The data were then processed using the XRTDAS software package (v.3.7.0) developed by the ASI Space Science Data Center\footnote{\url{https://www.ssdc.asi.it/}} (SSDC), released by the NASA High Energy Astrophysics Archive Research Center (HEASARC) in the HEASoft package (v.6.30.1). The data were reprocessed with the \texttt{xrtpipeline} script and using calibration files from \textit{Swift}-XRT CALDB (version 20210915). The X-ray spectrum was constructed from the calibrated and cleaned event file for each pointing. The events were selected within a radius of 20 pixels ($\sim$47 arcsec) in both WT and PC modes. The background was extracted from a nearby circular region with a radius of 40 pixels. The ancillary response files were generated with the \texttt{xrtmkarf} task applying the corrections for PSF losses and CCD defects using the cumulative exposure map. The $0.3-10$\,keV source spectra were binned using the \texttt{grppha} task by requiring at least 20 counts per energy bin. We used \texttt{XSPEC} with both a power-law and log-parabola model (with a pivot energy fixed at 1 keV and an added photoelectric absorption component assuming an equivalent hydrogen column density fixed to the Galactic value along the line of sight). In the overall majority of the observations, the preference for a log-parabola model is statistically significant (i.e., $>5\sigma$). The intrinsic fluxes were extracted in the 0.3-2\,keV, and 2-10\,keV energy bands.\par

Data from the \textit{Swift} Burst Alert Telescope (BAT) are publicly accessible online\footnote{\url{https://swift.gsfc.nasa.gov/results/transients/}}. In this work, the daily light curve is used from 15-50\,keV. \par

The Rossi X-ray Timing Explorer~\citep[\textit{RXTE};][]{1993A&AS...97..355B} satellite performed almost daily pointing observations of Mrk$\,$421 during the time period from December 12, 2009, (MJD 55177) to April 29, 2010, (MJD 55315). The data were analyzed using FTOOLS v6.9 following the settings and procedures recommended by the NASA RXTE Guest Observer Facility\footnote{\url{https://heasarc.gsfc.nasa.gov/docs/xte/xhp_proc_analysis.html}}. We produce a light curve from 2-10\,keV. For more details on the analysis settings see the instrument description given in~\citet{2015A&A...578A..22A}. \par

Data from the \textit{RXTE} All-Sky Monitor (ASM) are publicly accessible online\footnote{\url{http://xte.mit.edu/asmlc/ASM.html}}. In this work, the one-day average light curve in the energy range from 2-12\,keV is used. \par

\subsection{Ultraviolet}
\textit{Swift} also provides coverage in the ultraviolet (UV) band from the UV/Optical Telescope \citep[UVOT,][]{2005SSRv..120...95R}. We selected observations in the W1 (251\,nm), M2 (217\,nm), and W2 (188\,nm) filters by applying standard quality checks to all observations in the chosen time interval, excluding those with unstable attitudes or affected by contamination from a nearby starlight (51 UMa). For each individual observation, we performed photometry over the total exposures in each filter to extract flux values. The same apertures for source counts (the standard with 5\,arcsec radius) and background estimation (mostly three-four circles of $\sim$16\,arcsec radii off the source) were applied to all. We performed the photometry extraction with the official software included in the HEAsoft 6.23 package, from HEASARC, and then applied the official calibrations \citep{2011AIPC.1358..373B} included in the CALDB release (20201026). As a last step, the source fluxes were dereddened according to a mean interstellar extinction curve \citep{1999PASP..111...63F} and the mean Galactic $E(B - V)$ value of 0.0123 mag \citep{1998ApJ...500..525S, 2011ApJ...737..103S}. \par

\subsection{Optical}
Observations in the R-band were performed within the GLAST-AGILE Support Program (GASP) of the Whole Earth Blazar Telescope~\citep[WEBT; e.g.][]{2008A&A...481L..79V, 2009A&A...504L...9V}, including multiple optical telescopes around the globe. We use the data published in~\citet{2017MNRAS.472.3789C}, for which the contribution of the host galaxy has been subtracted. \par
Optical polarization measurements are included from the Lowell (Perkins), Crimean, Calar Alto, and Steward observatories, also taken from~\citet{2017MNRAS.472.3789C}. The contribution of the host galaxy is also taken into account for the polarization data. \par

\subsection{Radio}
The radio data were taken with the 14$\,$m Mets\"ahovi Radio Observatory at 37$\,$GHz, the 40$\,$m Owens Valley Radio Observatory (OVRO) telescope at 15$\,$GHz, and the 26$\,$m University of Michigan Radio Astronomy Observatory (UMRAO) at 14.5$\,$GHz. Observations at 225$\,$GHz (1.3$\,$mm) were performed at the Submillimeter Array (SMA) near the summit of Mauna Kea (Hawaii). For more details on the instruments and observations see the description given in~\citet{2015A&A...578A..22A}. \par
In addition, the Very Long Baseline Array (VLBA) performed observations at 43$\,$GHz. The data were provided by the Boston University blazar group\footnote{\url{https://www.bu.edu/blazars/VLBAproject.html}} and are part of their monitoring program of gamma-ray blazars. We obtained the images of the parsec-scale jet using the Astronomical Image Processing System (AIPS) and Difmap software packages~\citep{2017ApJ...846...98J}. \par

\section{Fitting the power spectral density} \label{sec:fit_PSD}
The PSD cannot be directly estimated from real data sets. Real observations suffer from unevenly sampled data with large gaps in the coverage. The limited time coverage as well as the temporal binning of the data can cause a transfer of variability power from lower to higher frequencies and vice versa. These distortions need to be accounted for obtaining an accurate estimate of the true PSD. \par
\begin{figure}[h]
\centering
  \resizebox{0.9\hsize}{!}{\includegraphics{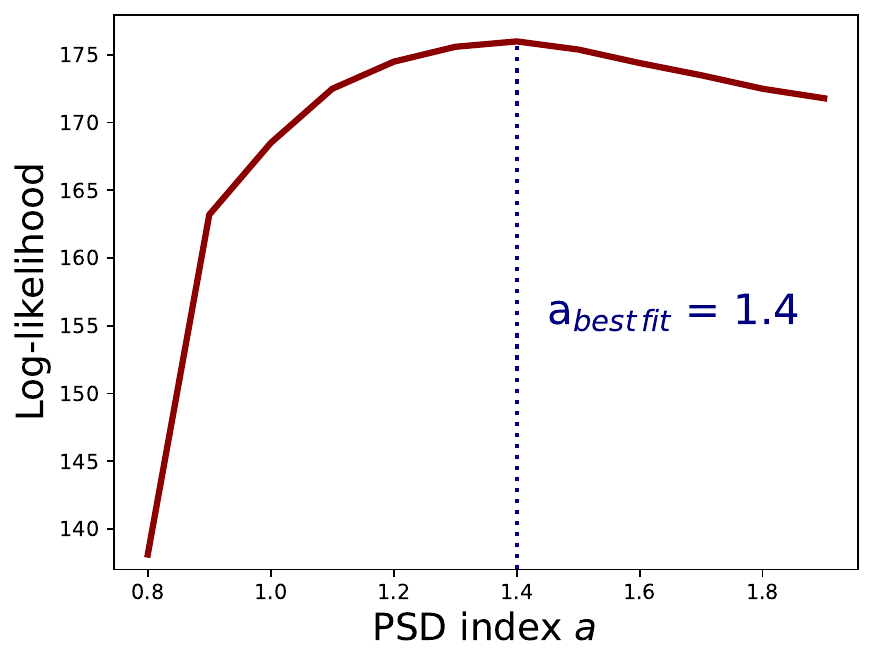}}
  \caption{Likelihood profile $\mathcal{L} (a) $ (see Eq.~\ref{eq:psd_likelihood}) for the 2-10\,keV band with its best fit value of 1.4 denoted by a vertical blue dotted line.}
  \label{fig:likelihood}
\end{figure}

We estimate the true PSD index using a simulation-based forward-folding procedure, using the code described in \citet{axel-thesis, 2021A&A...655A..89M}. We cover the parameter space from 0.3 to 2.5 using steps of 0.1. For each assumed spectral index, we simulate a set of 3000 light curves using the assumption of a power-law-shaped PSD. We use the underlying probability distribution described in~\citet{2013MNRAS.433..907E} to match the observed flux distribution as well as possible. We simulate light curves 10 times longer than the real light curves and apply the same temporal binning to account for the previously mentioned leakage effects. In order to compare the simulated with the real light curves, we use the multiple fractions variance function (MFVF) as a proxy for the PSD~\citep{2011A&A...531A.123K}. The MFVF is computed by splitting a given light curve in the temporal middle and taking the flux variance of both halves. The two halves are then split again into two, and the variances are estimated. This is repeated until a minimum time scale of one day is reached (which is the minimum temporal resolution of the light curves). The resulting variances as a function of time scale give a robust alternative representation of the PSD. The MFVF is computed for all simulated light curves, which results in a probability density function $p(a,f_i)$ for each out of $N$ frequency bins. It gives the probability of measuring a MFVF in the frequency bin $f_i$, assuming the underlying PSD with the index $a$. The best-fit value for the spectral index of the real light curve can then be estimated with a maximum likelihood estimation by finding the index $a$ maximizing
\begin{equation}
\label{eq:psd_likelihood}
\mathcal{L} (a) = \sum_{i=0}^{N} \text{ln} (p(a,f_i)) \, .
\end{equation}
Fig.~\ref{fig:likelihood} shows an example likelihood profile for the 2-10\,keV band. The resulting values for a selected set of light curves are given in Tab.~\ref{tab:PSD_Fit}.  \par
\begin{figure}[h]
\centering
  \resizebox{0.9\hsize}{!}{\includegraphics{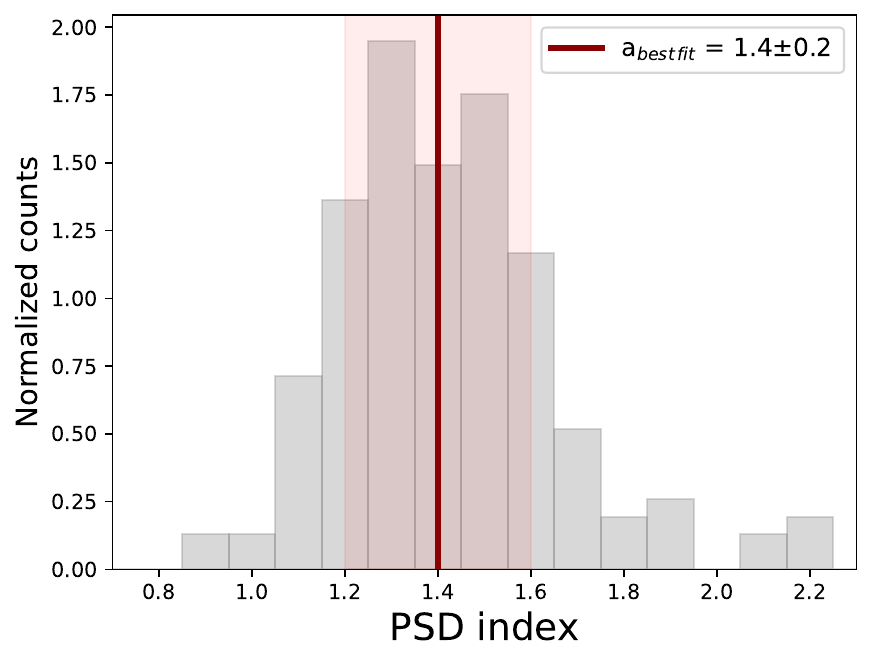}}
  \caption{Histogram of the best-fit indices derived from simulations using as input a simulated light curve that has a known PSD index $a=1.4$, in agreement with the real data for the 2-10\,keV band.}
  \label{fig:unc_hist}
\end{figure}

Following \citet{axel-thesis, 2021A&A...655A..89M} we estimated the uncertainties of the indices by simulating 100 light curves using the best-fit index for each light curve and then refit all 100 created light curves with the same method. This creates a histogram of the resulting indices, where the uncertainties are given by the 68\% (1$\,\sigma$) containment region. As an example, Fig.~\ref{fig:unc_hist} shows the resulting histogram for the 2-10\,keV band. Since the histograms can show a skewed distribution, we separately provide upper and lower uncertainties. \par

\section{Intranight VHE variability} \label{sec:appendix_intranight}
\begin{figure}[ht]
\centering
  \resizebox{\hsize}{!}{\includegraphics{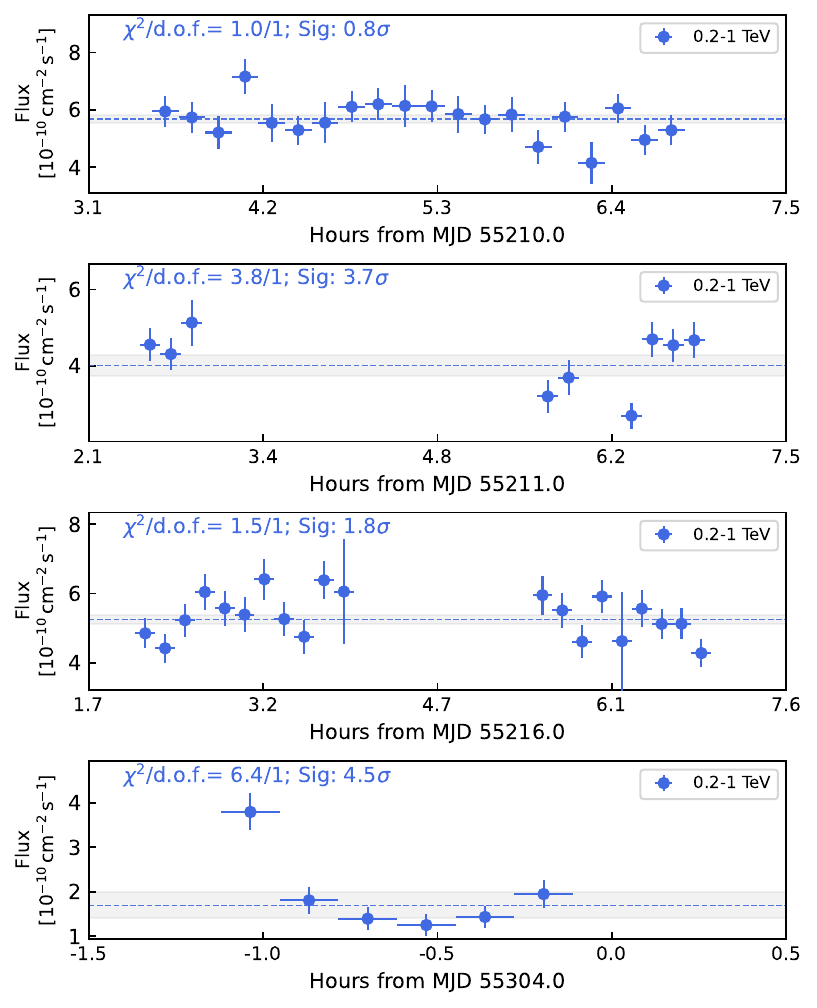}}
  \caption{Intranight VHE variability in the two energy bands 0.2-1\,TeV and >1\,TeV. The data are binned in 10\,min. The dotted line shows the constant fits for the 0.2-1\,TeV band.}
  \label{fig:intranight}
\end{figure}

Fig.~\ref{fig:intranight} shows the intranight light curves in the 0.2-1\,TeV band of four selected nights in 10\,min bins. The upper panel shows the January 14, 2010, (MJD 55210) light curve, where the highest fluxes are recorded by MAGIC during the flare in January. We fit the data with a constant model in order to test the hypothesis of a non-variable emission. The source shows no significant variability but a stable and high emission throughout the full exposure of around 3 hours. \par
In the second panel, the following night of January 15 (MJD 55211) is shown. The hypothesis of a constant emission is rejected at 3.7\,$\sigma$, indicating a significant variability on a time-scale of well below 1 hour in VHE gamma rays. Missing data points in between were caused by technical interruptions. \par
During the flare in January, the VHE flux increases again around January 20 (MJD 55216). We see no significant variability but again a strong but constant emission throughout the exposure. \par
Lastly, we show April 18 (MJD 55304) in the lowest panel. We observe a quick decay and small rise of the flux, which corresponds to significant variability at 4.5\,$\sigma$. \par

\section{Hardness ratios of VHE gamma rays and X-rays} \label{sec:appendix_HR}
\begin{figure}[ht]
\centering
  \resizebox{\hsize}{!}{\includegraphics{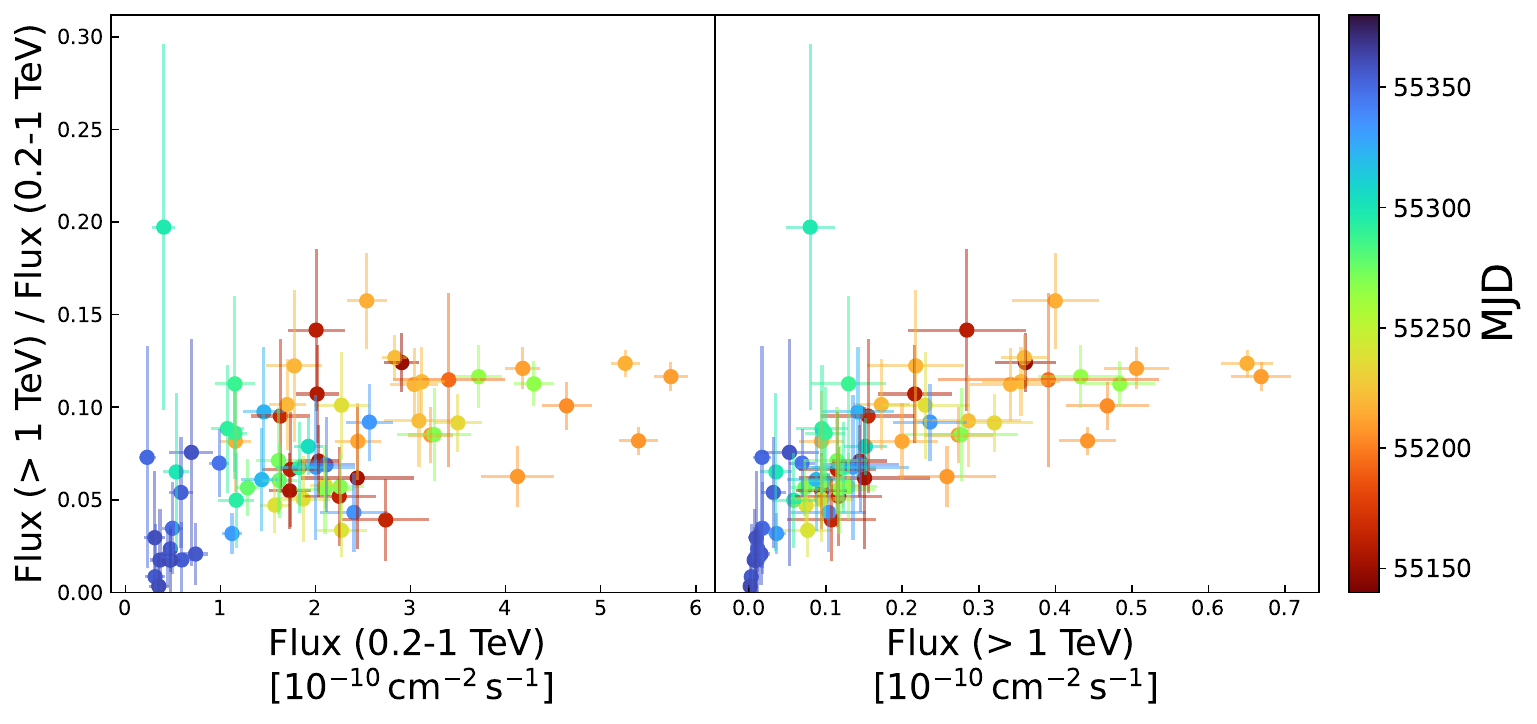}}
  \caption{Hardness ratios as a function of the flux 0.2-1\,TeV  (left) and above 1\,TeV(right) obtained by MAGIC. The color indicates the time of the observation in MJD.}
  \label{fig:HR_magic}
\end{figure}

Fig.~\ref{fig:HR_magic} shows the hardness ratios (HR) of all MAGIC observations as a function of the flux between 0.2-1\,TeV in the left plot and above 1\,TeV in the right one. Both plots show a clear harder-when-brighter trend. The rise of the HR with the flux is steep for low fluxes but flattens at higher flux levels. In the right plot, the HR seems to remain constant with a rising flux above $\sim$$3 \times 10^{-11}\,\text{cm}^{-2}\text{s}^{-1}$. The plot on the left shows similar behavior, but the overall scatter of the data is slightly higher. This flattening harder-when brighter-trend at VHE gamma rays is consistent with previous reports~\citep[e.g.][]{2021MNRAS.504.1427A}. \par 

\begin{figure}[ht]
\centering
  \resizebox{\hsize}{!}{\includegraphics{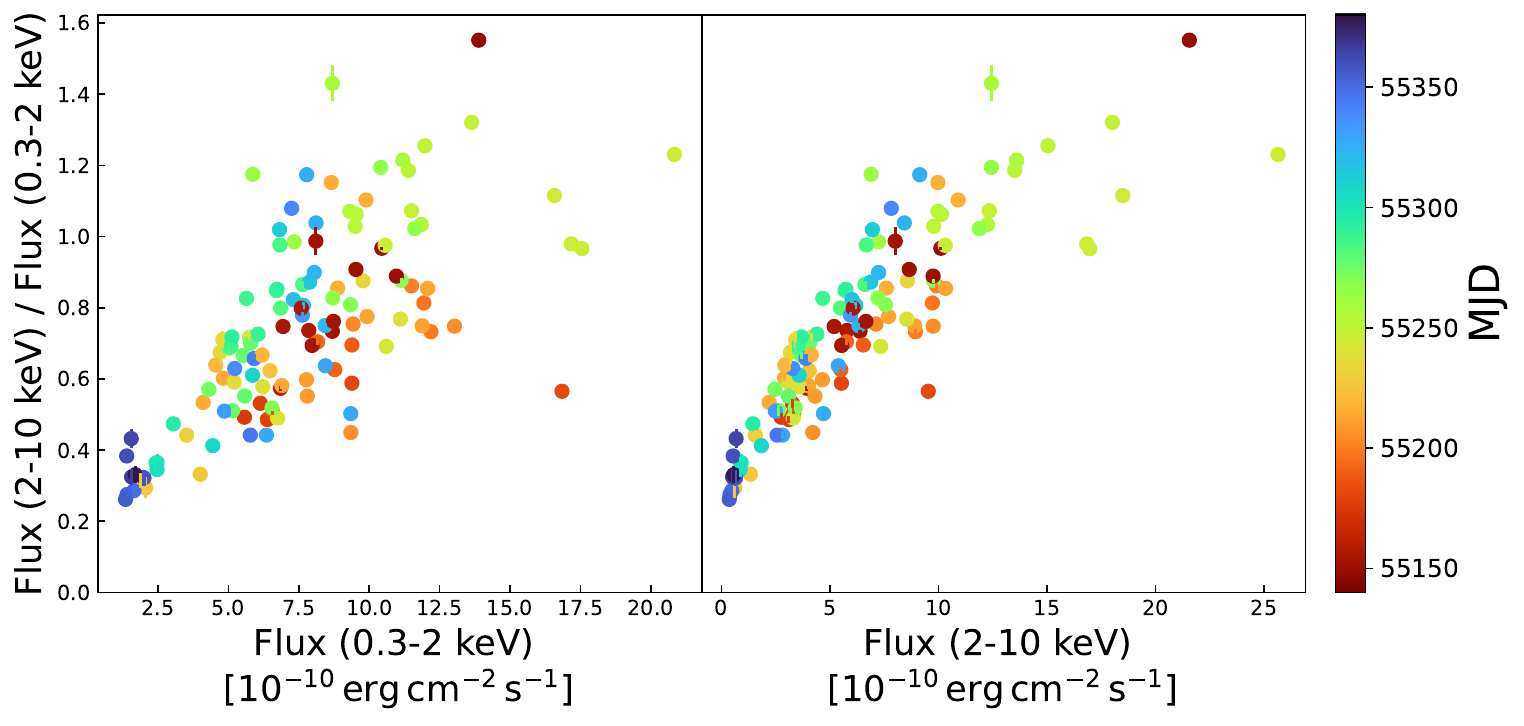}}
  \caption{Hardness ratios as a function of the flux between 0.3-2\,keV (left) and 2-10\,keV (right) obtained by \textit{Swift}-XRT. The color indicates the time of the observation in MJD.}
  \label{fig:HR_XRT}
\end{figure}

Fig.~\ref{fig:HR_XRT} shows the HRs of the \textit{Swift}-XRT observations as a function of the fluxes between 0.3-2\,keV in the left and 2-10\,keV in the right plot. At the end of the campaign, when the source is in a low state of activity, the HR goes as low as $\sim$0.26. Similar to the VHE data, a clear harder-when-brighter trend is visible. On the contrary, the HR does not show a clear flattening, but a rather linear behavior. This is especially pronounced in the right plot. In the left plot for the lower-energy fluxes, the data is more scattered. The highest HR of $\sim$1.55, directly at the start of the campaign, still seems to follow an almost linear trend. The HR ratio corresponding to the highest flux values, however, is more compatible with a flattening of the trend. Overall, the flattening of the HR is much less pronounced than in the results from~\citet{2021MNRAS.504.1427A}. \par

\section{Estimating the significances of correlations} \label{sec:significances_scatter}
We simulate a set of 100.000 uncorrelated pairs of light curves for each combination of energy bands, following the previous prescription, in Appendix~\ref{sec:fit_PSD}. To get the same degree of variability at different frequencies as the real data, we use the PSDs obtained in Sec.~\ref{sec:PSD}. The simulated light curves are generated with a temporal precision matching the typical observation time of the observations and then binned with the same temporal sampling as the real data. We then compute the coefficient for each pair. The resulting distributions are fitted with a Gaussian Kernel model to approximate the probability density function (PDF). Integration of the PDF above the coefficient of the real dataset provides a p-value. It indicates the probability of finding uncorrelated datasets that have a correlation at least as extreme as the one computed from the real dataset. We then translate the p-value into a significance expressed in levels of 1\,$\sigma$. It must be emphasized that the standard approach for assessing the significance of the Pearson coefficient under the assumption of two Gaussian-distributed data sets is not applicable. The given datasets show flux distributions with strong tails towards higher fluxes. Additionally, measurement uncertainties are not taken into account in the standard method. We, therefore, rely on simulations to estimate the significance, which gives a more robust and conservative estimate. \par 

Besides the Pearson correlation, many studies also make use of the discrete correlation function~\citep[DCF;][]{1988ApJ...333..646E} to quantify the correlation between fluxes in two energy bands. However, in this particular case of correlating the X-ray and the VHE gamma-ray light curves, we found that the DCF approach is less sensitive and yields somewhat underestimated values for the correlation strength. The resulting underestimation of the correlation occurs because the VHE and X-ray light curves have very different temporal coverage, with the X-rays having a much denser sampling with substantially fewer gaps. This is visible, for instance, between December 2009 and mid-January 2010, or between mid-February 2010 and early March 2010 (see Fig.~\ref{fig:MWL_LC}), where no VHE gamma-ray observations are available, while the densely \textit{Swift}-XRT light curve unveils strong variability. Since the normalization of the DCF depends on the standard deviation from the entire light curve and not only from the VHE/X-ray simultaneous measurements (as is the case for the Pearson coefficient), the DCF strategy, for this specific dataset is biased towards lower values, given the strong X-ray variations in periods without a VHE coverage. The mismatch in the coverage between the X-ray and VHE light curves also generates a larger spread in the DCF of the simulated light curves (used to determine its significance), further reducing the significance of the measured correlations. These effects were already reported in \citet{2014MNRAS.445..437M}. Therefore, we concluded that, for this specific case of the X-ray and VHE gamma-ray light curves, the DCF strategy is not adequate to properly quantify the magnitude of the correlation and its related significance. On the other hand, the local cross-correlation function \citep[LCCF;][]{1999PASP..111.1347W} provides an alternative approach to that of the DCF, since it also takes into account the measurement uncertainties, but differently to the DCF, the values are normalized with the standard deviation from only the coincident measurements (analogous to the Pearson coefficient). As a cross-check, we evaluated all correlations estimated from the flux-flux plots in this work with the LCCF, and we found that the values and significances are almost identical to those obtained for the Pearson coefficient.

\section{Discrete correlation functions} \label{sec:appendix_corr_noflare}
This section presents the results of the DCF analysis referenced in the main text. \par

\begin{figure*}
     \centering
     \begin{subfigure}[t]{0.45\textwidth}
         \centering
         \includegraphics[width=\textwidth]{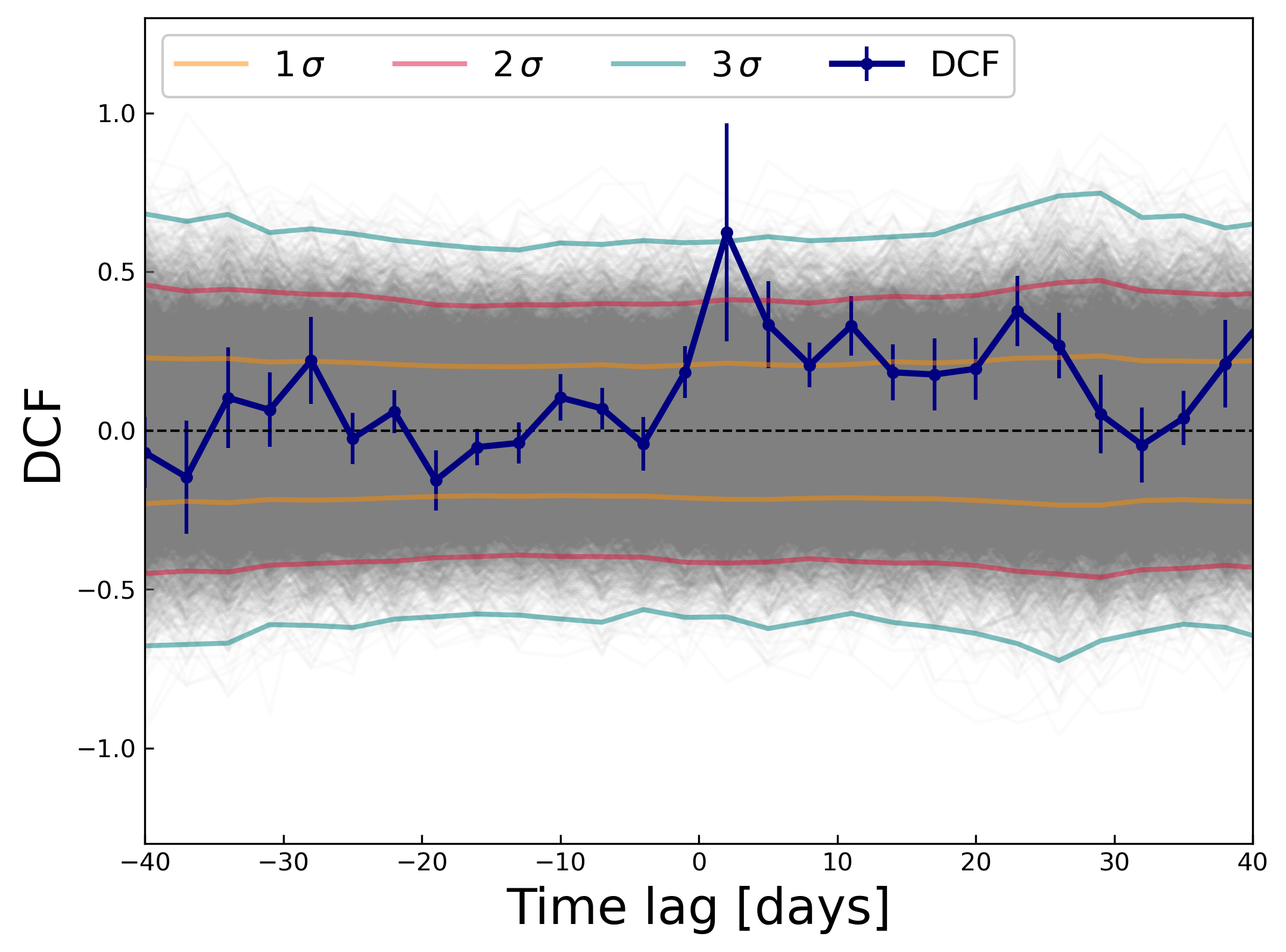}
         \caption{}
         \label{fig:VHE_vs_fermi_HE}
     \end{subfigure}
     \hspace{15pt}
     \begin{subfigure}[t]{0.45\textwidth}
         \centering
         \includegraphics[width=\textwidth]{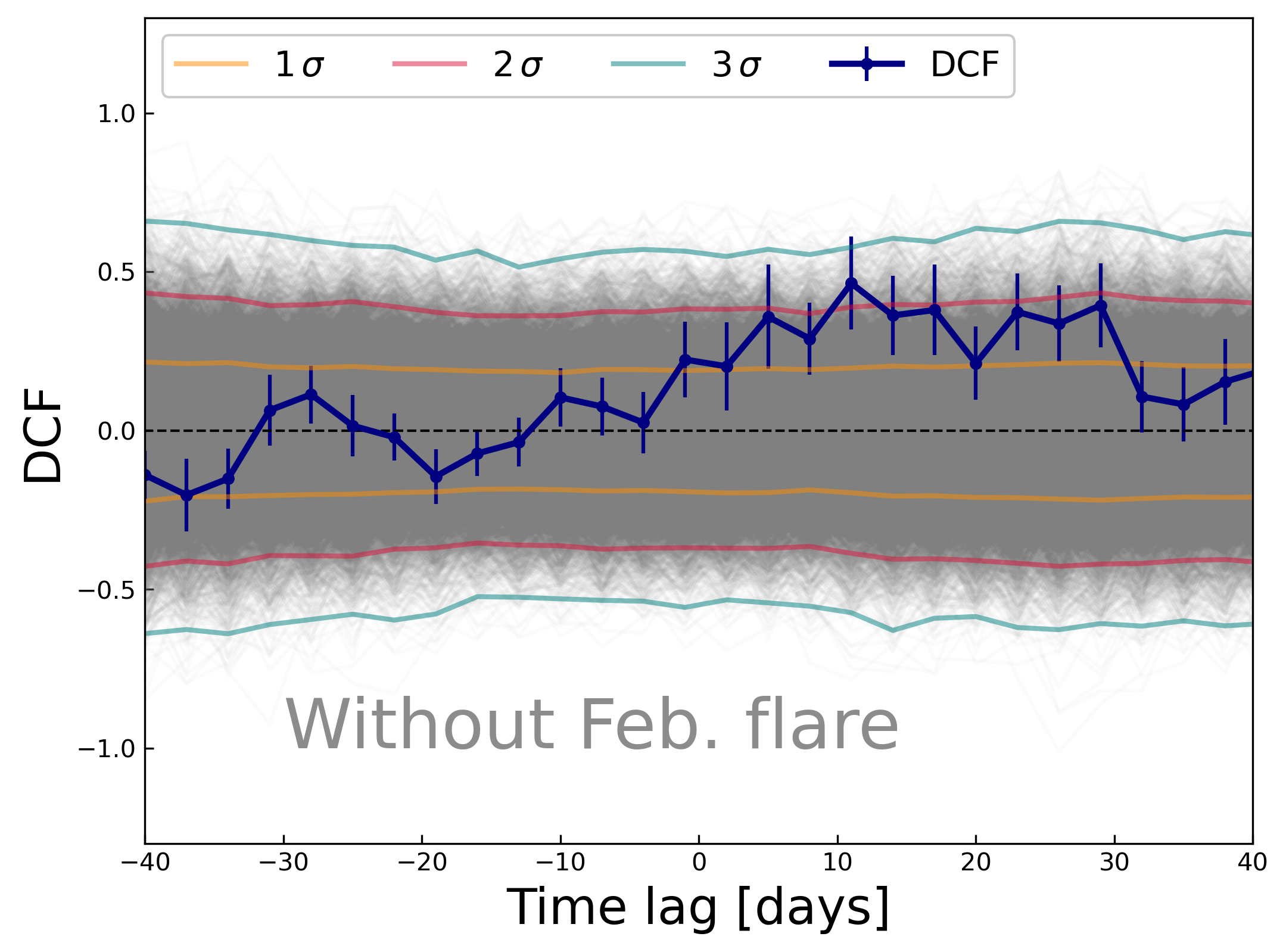}
         \caption{}
         \label{fig:VHE_vs_fermi_HE_noflare}
     \end{subfigure}
     
     \begin{subfigure}[t]{0.45\textwidth}
         \centering
         \includegraphics[width=\textwidth]{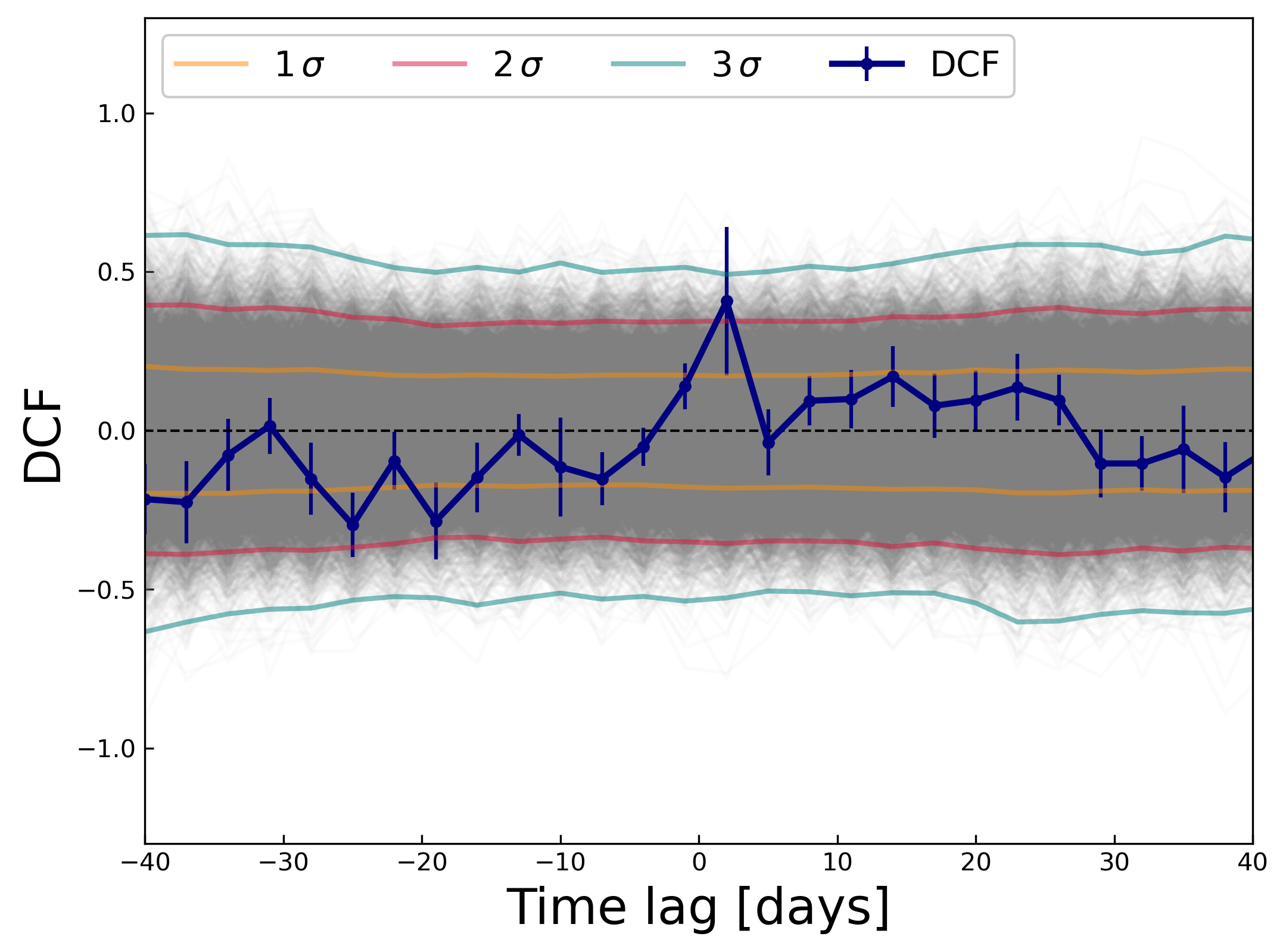}
         \caption{}
         \label{fig:VHE_vs_Fermi_LE}
      \end{subfigure}
     \hspace{15pt}
     \begin{subfigure}[t]{0.45\textwidth}
         \centering
         \includegraphics[width=\textwidth]{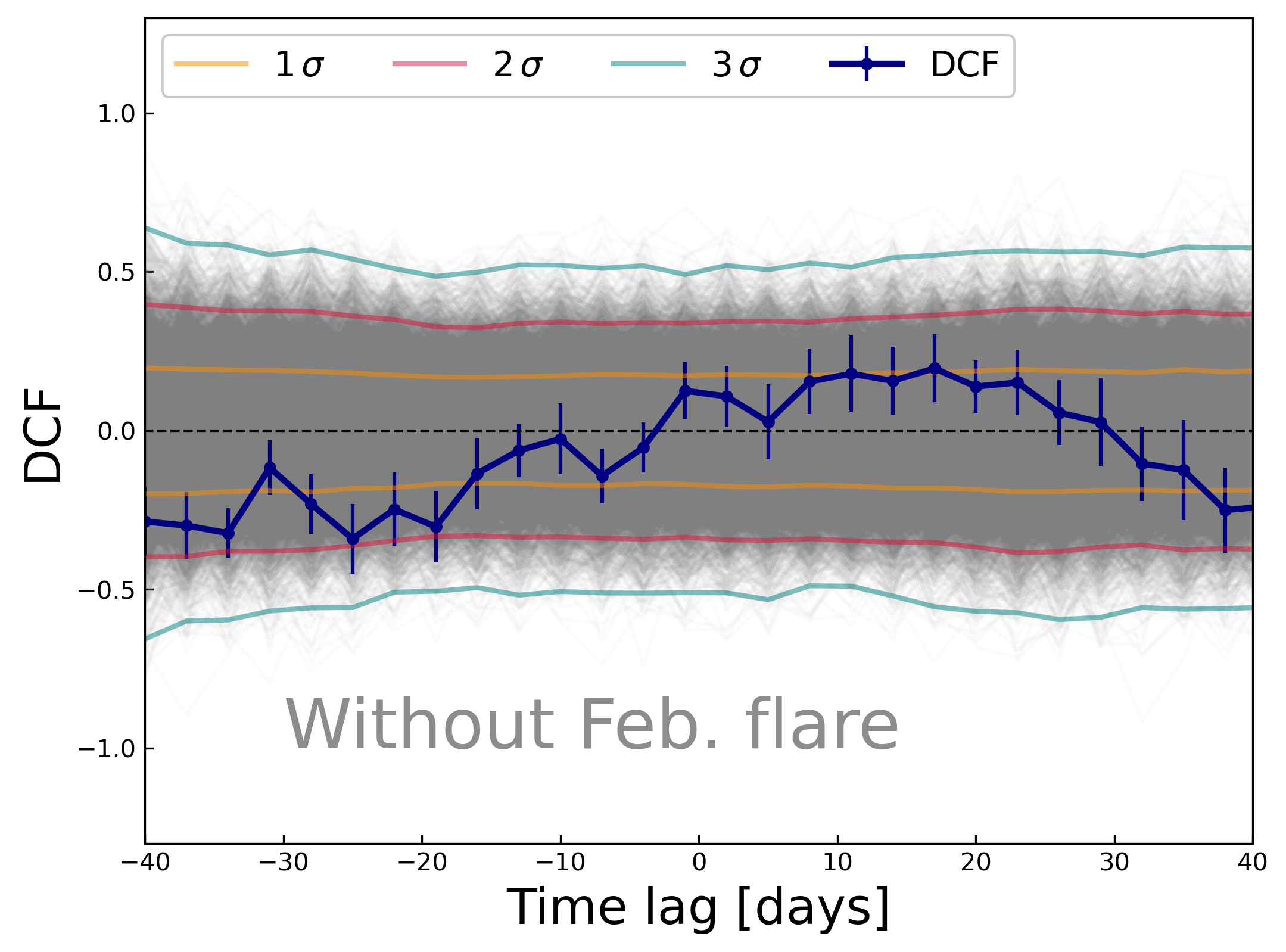}
         \caption{}
         \label{fig:VHE_vs_Fermi_LE_noflare}
     \end{subfigure}
    \caption{Discrete correlation function computed between the VHE gamma-ray fluxes above 0.2\,TeV, as measured by MAGIC and VERITAS, and the HE gamma-ray fluxes measured by \textit{Fermi}-LAT. The DCF is computed using a time bin of 3-days for a range of time lags between -40 to +40 days. The 1\,$\sigma$, 2\,$\sigma$, and 3\,$\sigma$ confidence levels obtained by simulations are shown by the yellow, red, and green lines, respectively. (a) >0.2\,TeV versus 3-300\,GeV; (b) >0.2\,TeV versus 3-300\,GeV without the flare in February 2010; (c) >0.2 TeV versus 0.3-3\,GeV; (d) >0.2 TeV versus 0.3-3\,GeV without the flare in February 2010.} 
        \label{fig:VHE_vs_Fermi}
\end{figure*}

\begin{figure*}
    \centering  
     \begin{subfigure}[t]{0.35\textwidth}
         \centering
         \includegraphics[width=\textwidth]{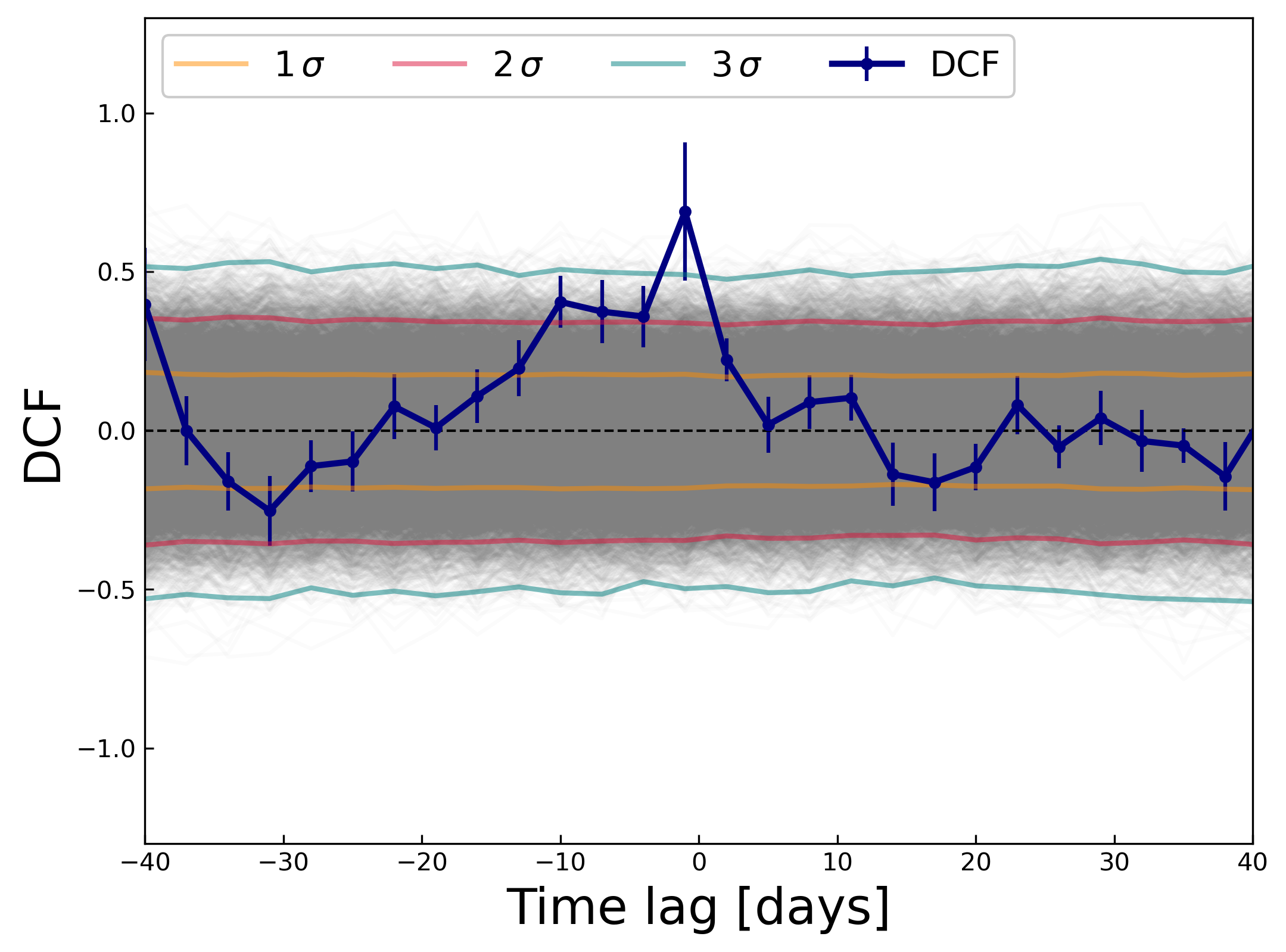}
         \caption{}
         \label{fig:Fermi_HE_vs_Xray_LE}
     \end{subfigure}  
     \hspace{15pt}
     \begin{subfigure}[t]{0.35\textwidth}
         \centering
         \includegraphics[width=\textwidth]{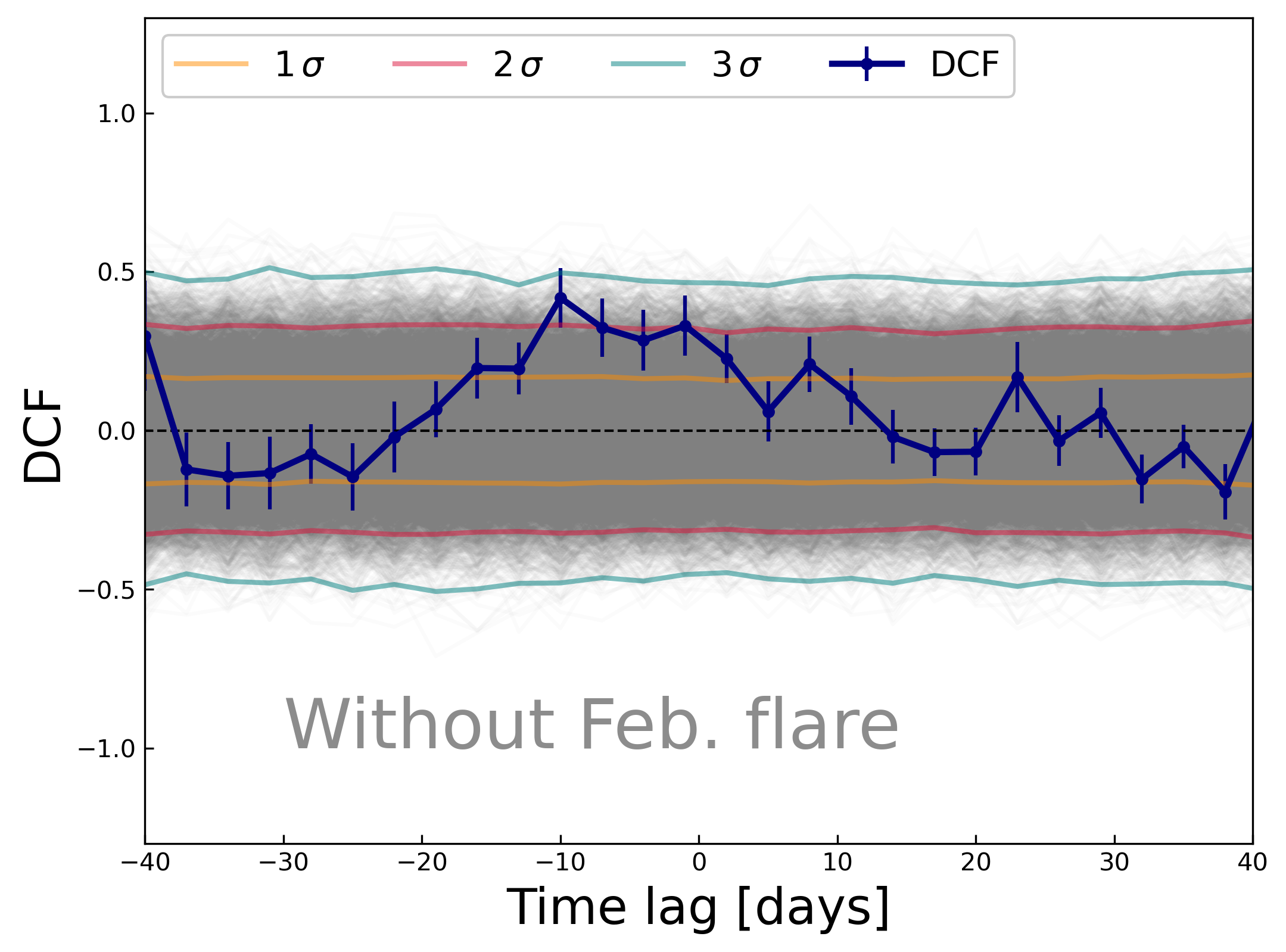}
         \caption{}
         \label{fig:Fermi_HE_vs_Xray_LE_noflare}
     \end{subfigure}    
     
     \begin{subfigure}[t]{0.35\textwidth}
         \centering
         \includegraphics[width=\textwidth]{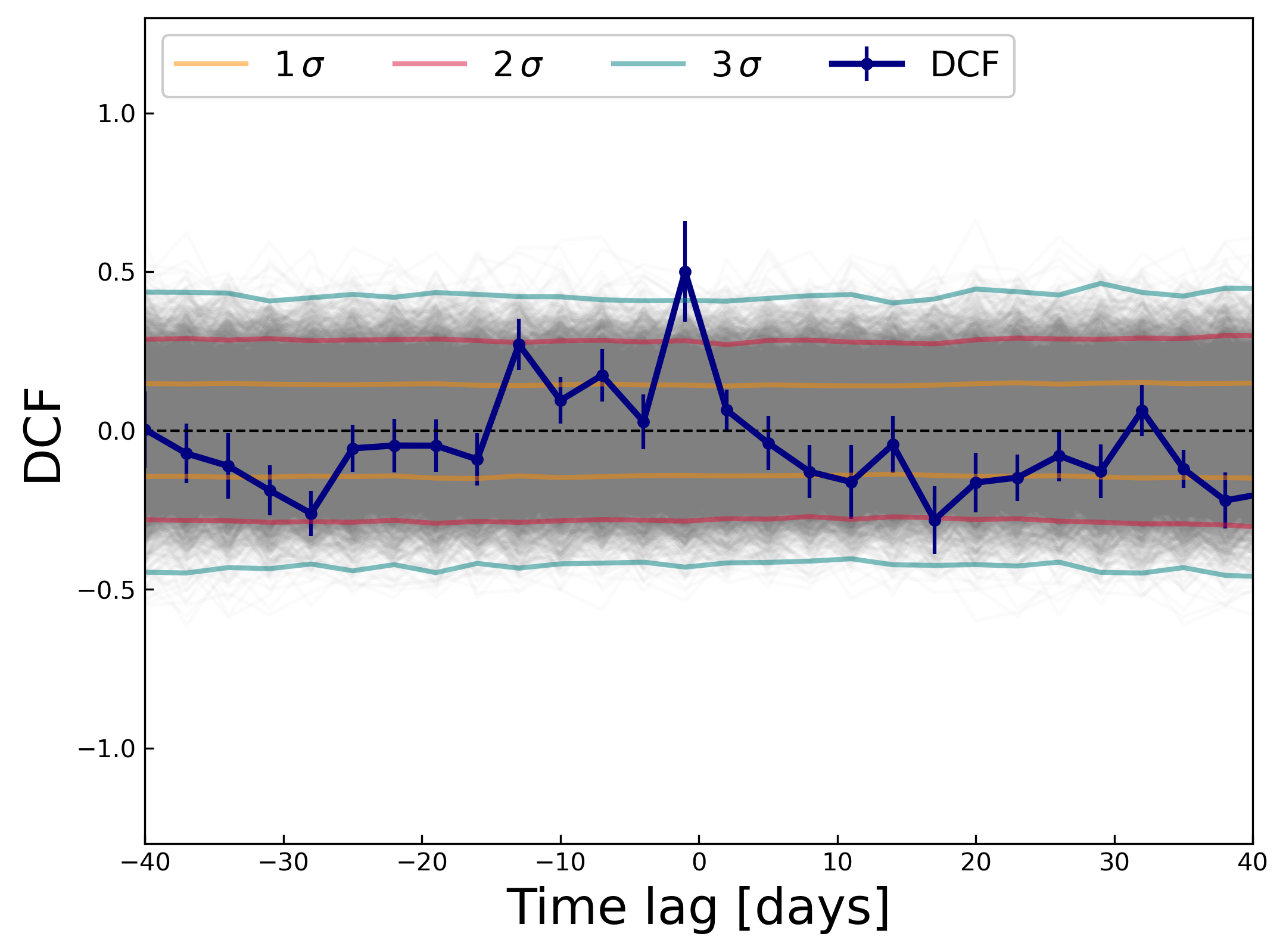}
         \caption{}
         \label{fig:Fermi_LE_vs_Xray_LE}
     \end{subfigure}
     \hspace{15pt}
     \begin{subfigure}[t]{0.35\textwidth}
         \centering
         \includegraphics[width=\textwidth]{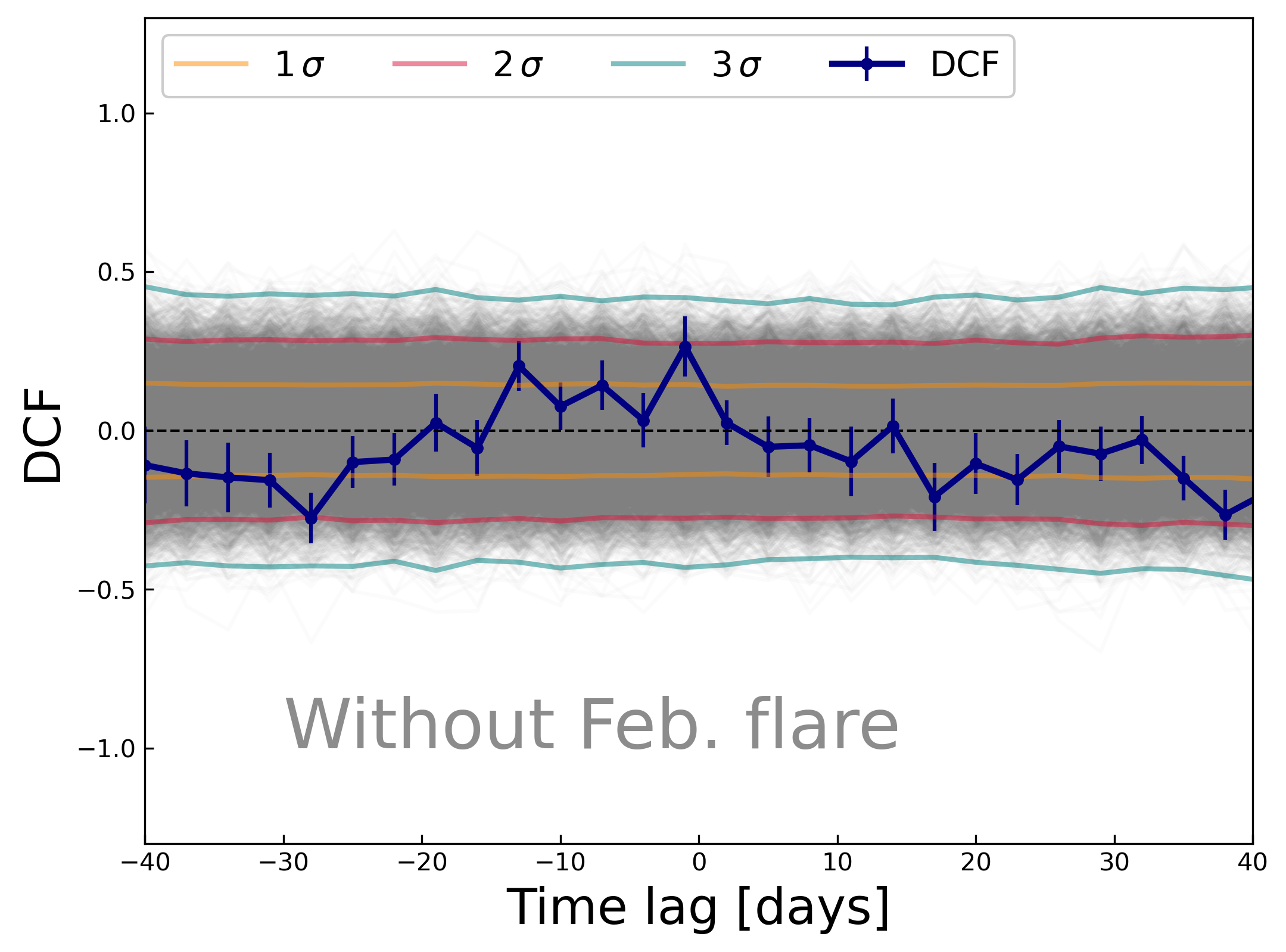}
         \caption{}
         \label{fig:Fermi_LE_vs_Xray_LE_noflare}
     \end{subfigure} 
    
     \begin{subfigure}[t]{0.35\textwidth}
         \centering
         \includegraphics[width=\textwidth]{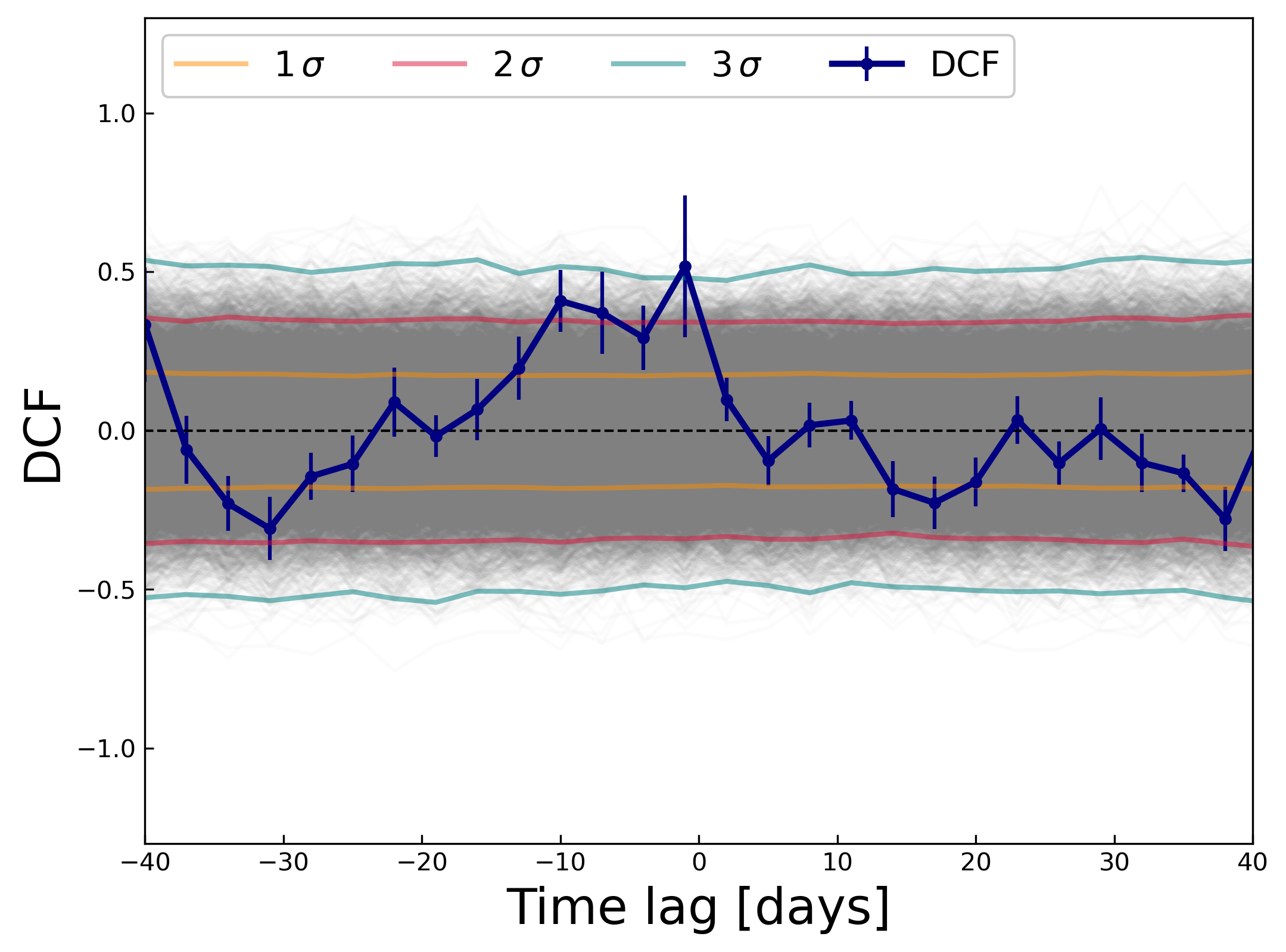}
         \caption{}
         \label{fig:Fermi_HE_vs_Xray_HE}
     \end{subfigure}  
     \hspace{15pt}
    \begin{subfigure}[t]{0.35\textwidth}
         \centering
         \includegraphics[width=\textwidth]{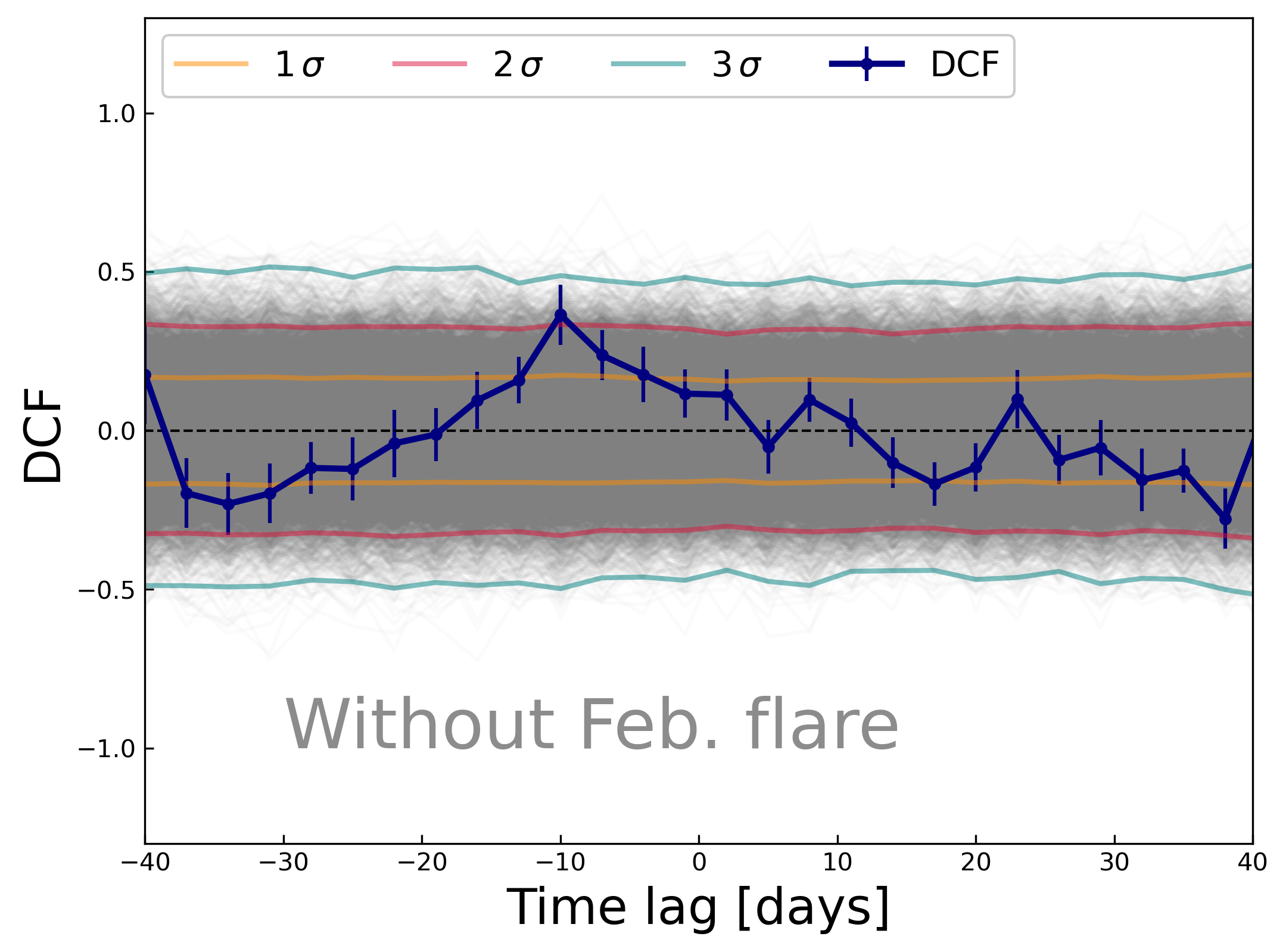}
         \caption{}
         \label{fig:Fermi_HE_vs_Xray_HE_noflare}
     \end{subfigure}   
     
     \begin{subfigure}[t]{0.35\textwidth}
         \centering
         \includegraphics[width=\textwidth]{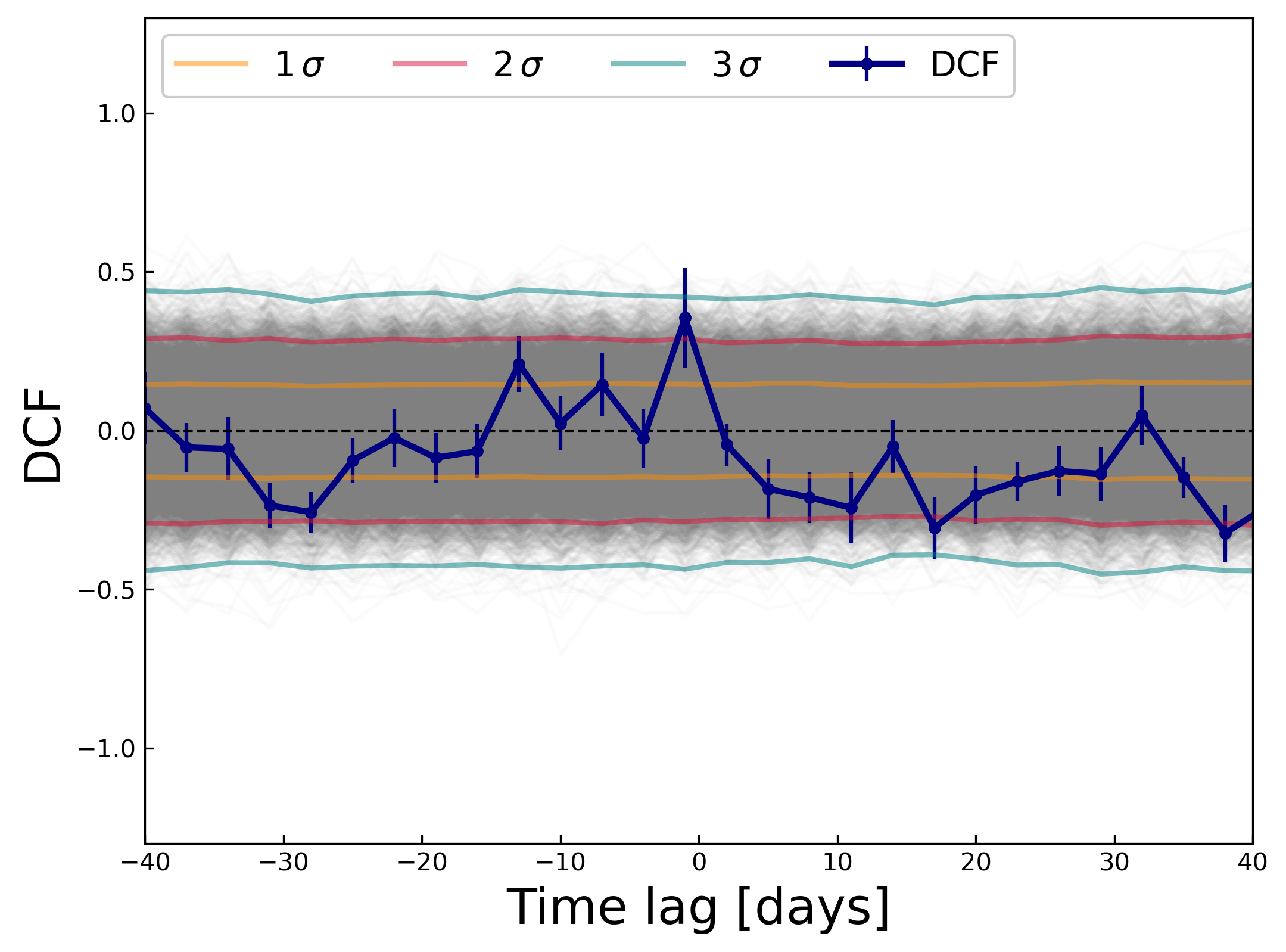}
         \caption{}
         \label{fig:Fermi_LE_vs_Xray_HE}
     \end{subfigure}
     \hspace{15pt}
     \begin{subfigure}[t]{0.35\textwidth}
         \centering
         \includegraphics[width=\textwidth]{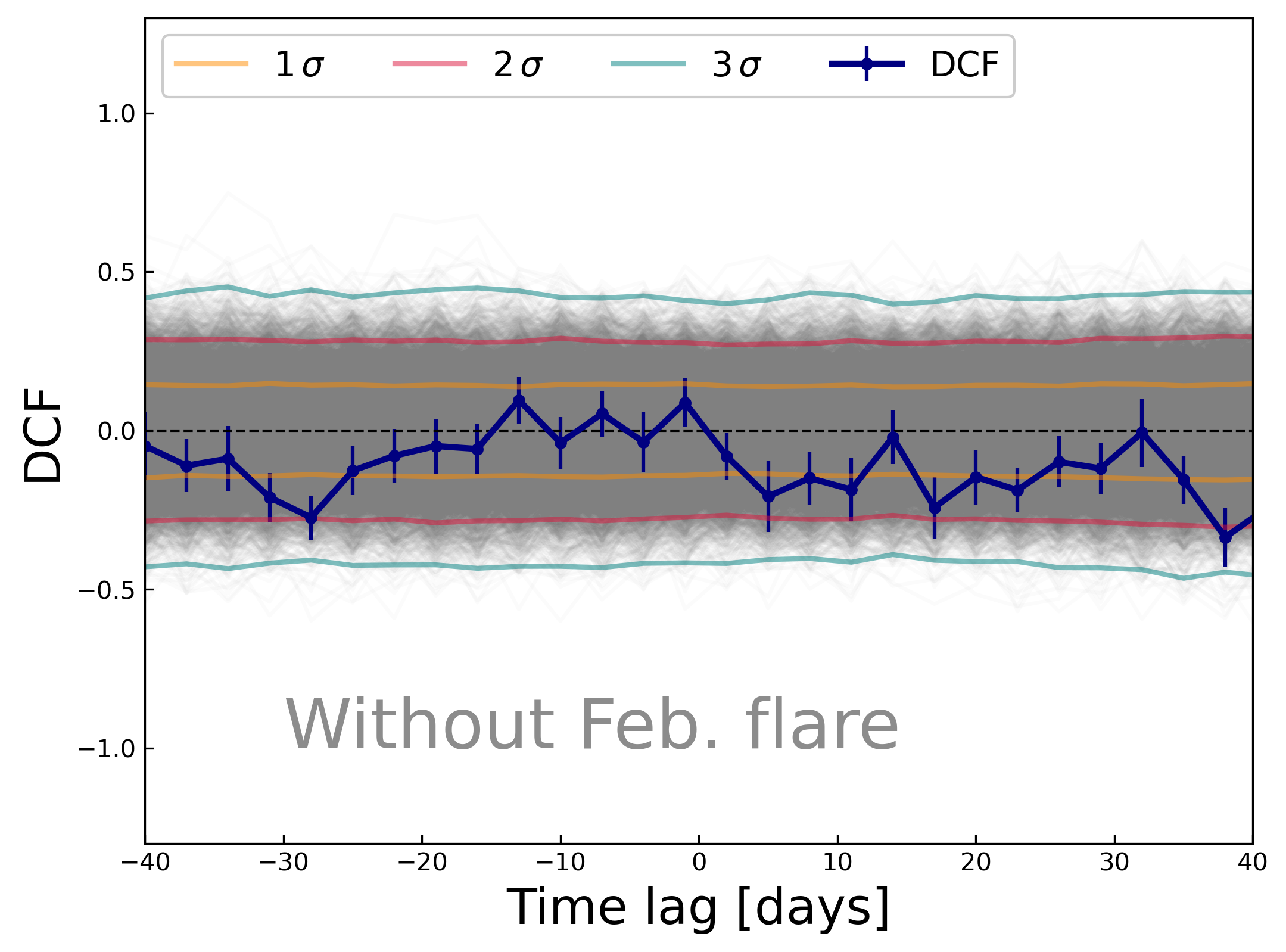}
         \caption{}
         \label{fig:Fermi_LE_vs_Xray_HE_noflare}
     \end{subfigure} 
     
        \caption{Discrete correlation function computed between two energy ranges provided by \textit{Fermi}-LAT and \textit{Swift}-XRT with and without the big flare in February using a binning of 3 days. It is computed for a range of time lags between -40 to +40 days. The 1\,$\sigma$, 2\,$\sigma$, and 3\,$\sigma$ confidence levels obtained by simulations are shown by the yellow, red, and green lines, respectively. (a) 3-300\,GeV versus 0.3-2\,keV; (b) 3-300\,GeV versus 0.3-2\,keV without the flare in February 2010; (c) 0.3-3\,GeV versus 0.3-2\,keV; (d) 0.3-3\,GeV versus 0.3-2\,keV without the flare in February 2010; (e) 3-300\,GeV versus 2-10\,keV; (f) 3-300\,GeV versus 2-10\,keV without the flare in February 2010; (g) 0.3-3\,GeV versus 2-10\,keV; (h) 0.3-3\,GeV versus 2-10\,keV without the flare in February 2010.}
        \label{fig:Fermi_Xray}
\end{figure*}

\begin{figure*}
     \begin{subfigure}[t]{0.45\textwidth}
         \centering
         \includegraphics[width=\textwidth]{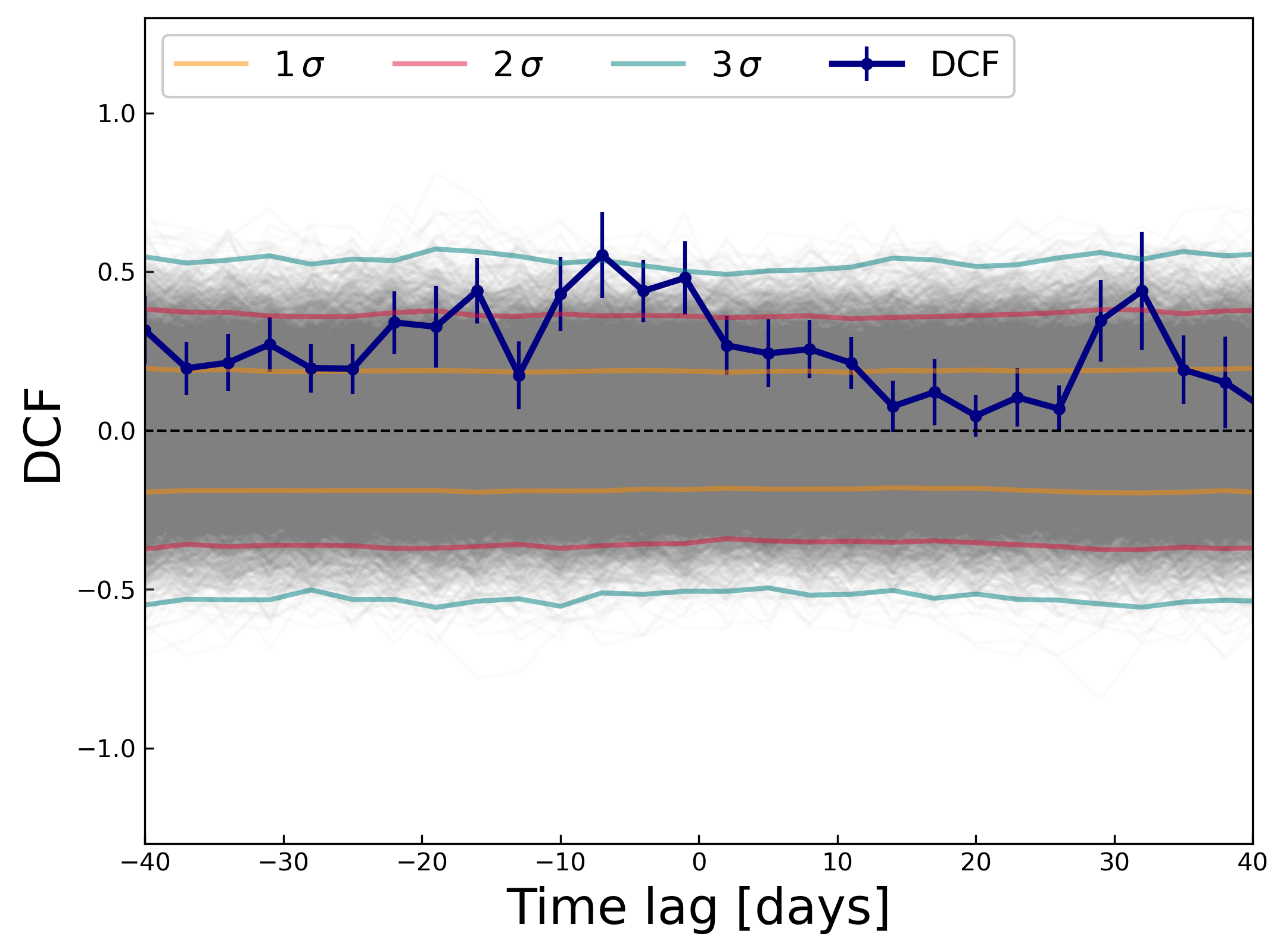}
         \caption{}
         \label{fig:Fermi_HE_vs_UV}
     \end{subfigure}
     \hspace{15pt}
     \begin{subfigure}[t]{0.45\textwidth}
         \centering
         \includegraphics[width=\textwidth]{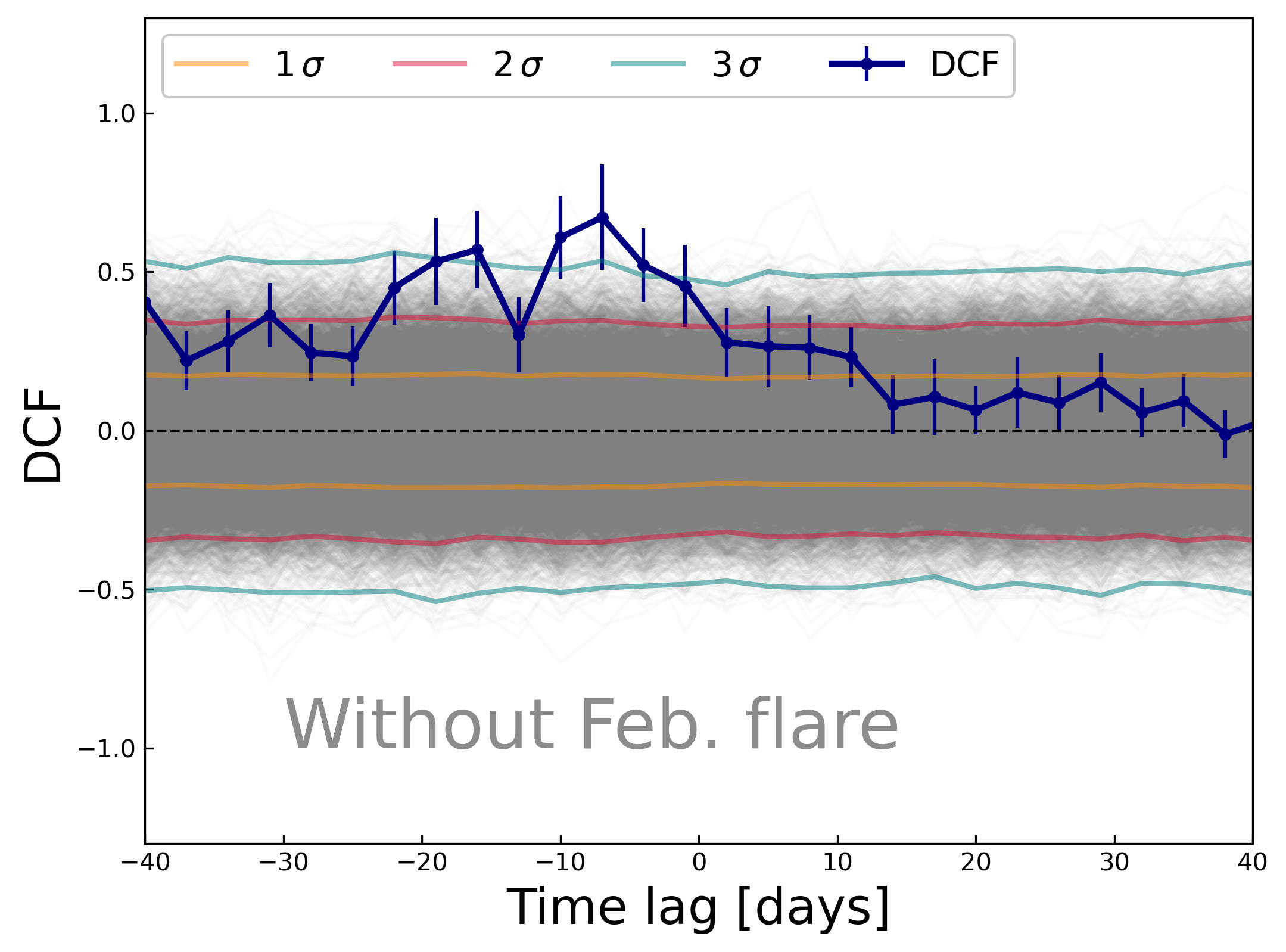}
         \caption{}
         \label{fig:Fermi_HE_vs_UV_noflare}
        \vspace{10pt}    
     \end{subfigure}
     
     \begin{subfigure}[t]{0.45\textwidth}
         \centering
         \includegraphics[width=\textwidth]{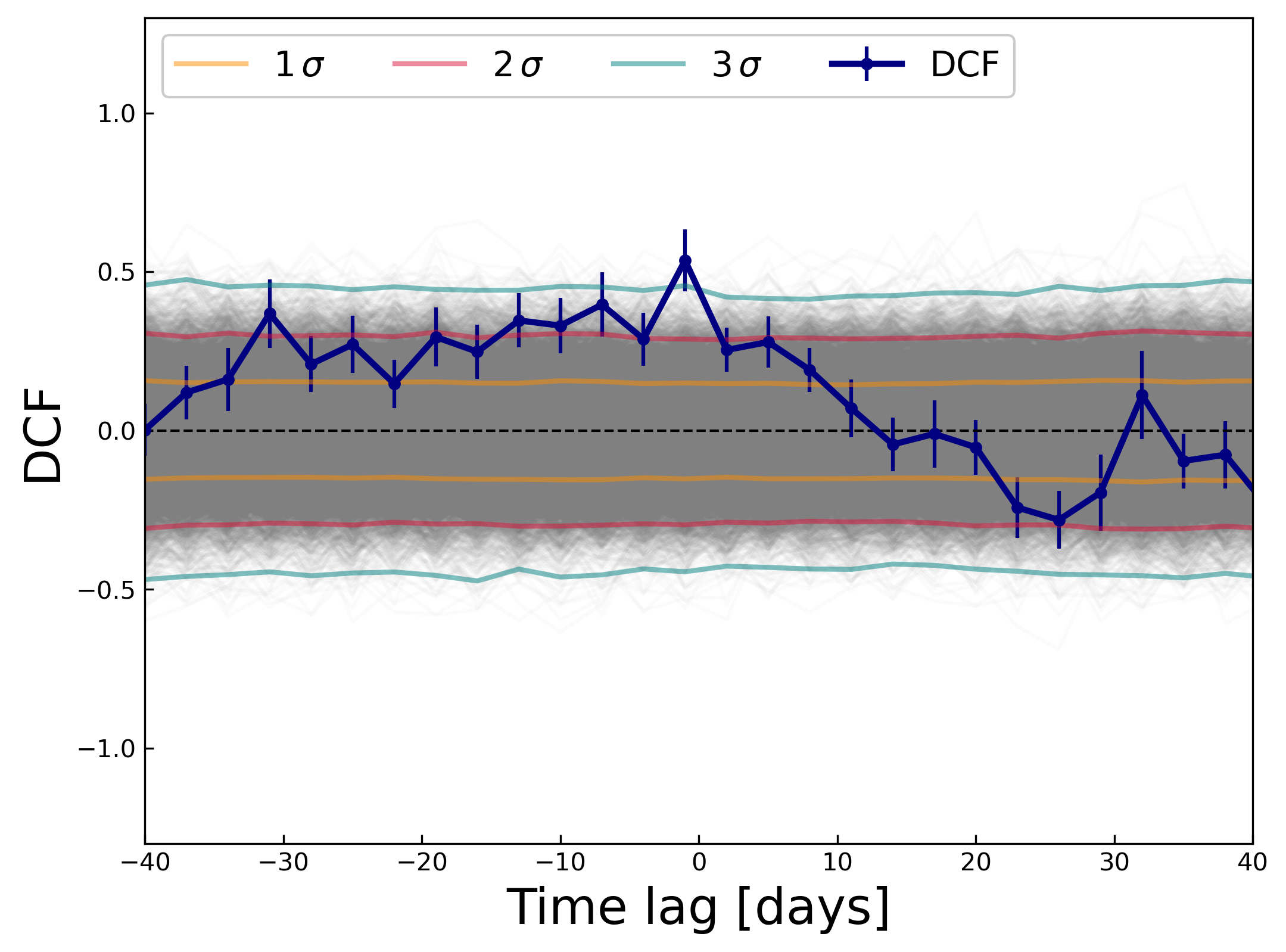}
         \caption{}
         \label{fig:Fermi_LE_vs_UV}
     \end{subfigure}
     \hspace{15pt}
     \begin{subfigure}[t]{0.45\textwidth}
         \centering
         \includegraphics[width=\textwidth]{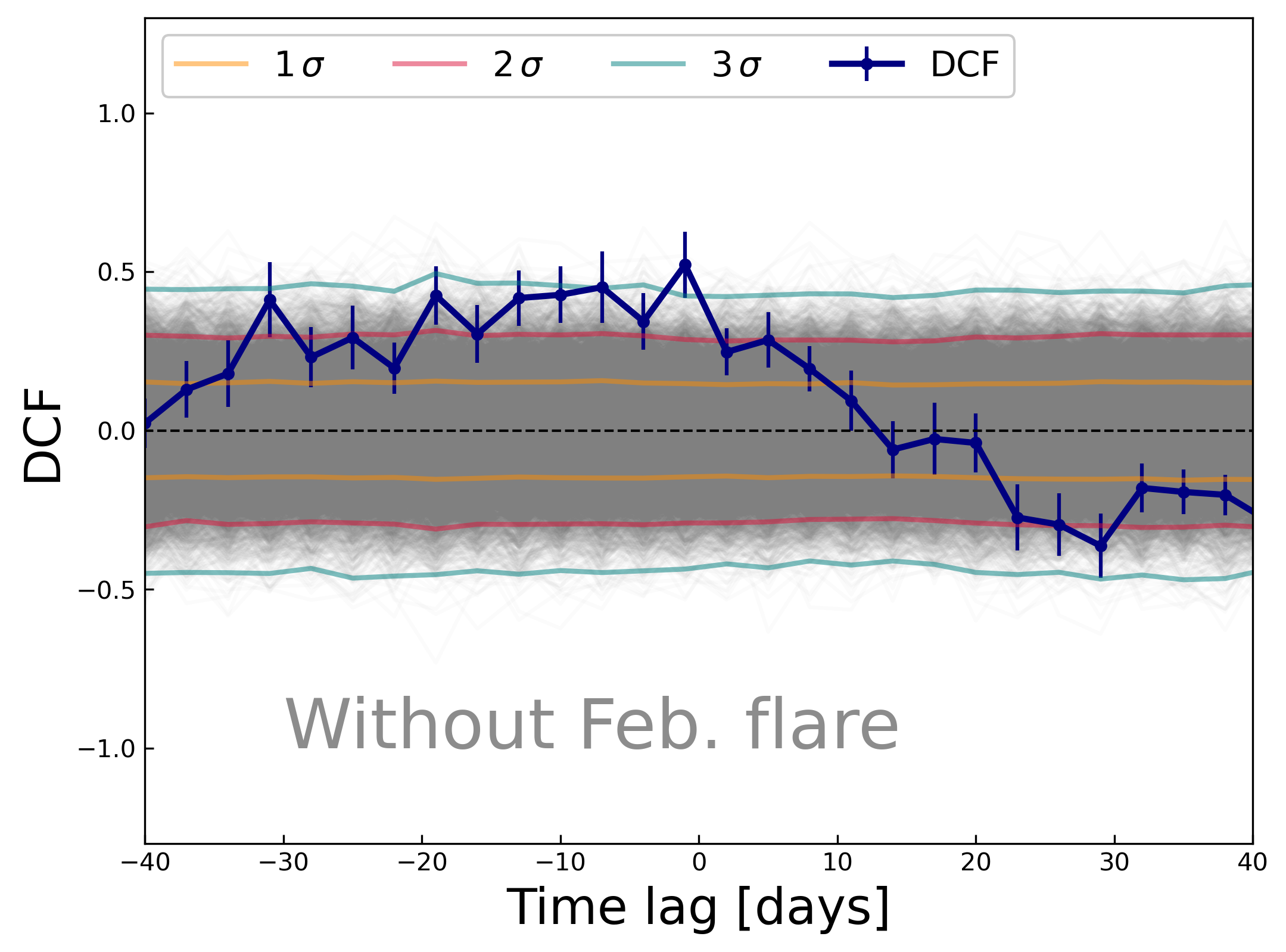}
         \caption{}
         \label{fig:Fermi_LE_vs_UV_noflare}
        \vspace{10pt}    
     \end{subfigure}
    \caption{Discrete correlation function computed between the two energy ranges provided by \textit{Fermi}-LAT and the W1 filter by \textit{Swift}-UVOT without the big flare in February using a binning of 3 days. It is computed for a range of time lags between -40 to +40 days. The 1\,$\sigma$, 2\,$\sigma$, and 3\,$\sigma$ confidence levels obtained by simulations are shown by the yellow, red, and green lines, respectively. (a) 3-300\,GeV versus W1; (b) 3-300\,GeV versus W1 without the flare in February 2010; (c) 0.3-3\,GeV versus W1; (d) 0.3-3\,GeV versus W1 without the flare in February 2010.}
        \label{fig:Fermi_vs_UV}
\end{figure*}

\begin{figure*}

     \begin{subfigure}[t]{0.45\textwidth}
     \centering
         \includegraphics[width=\textwidth]{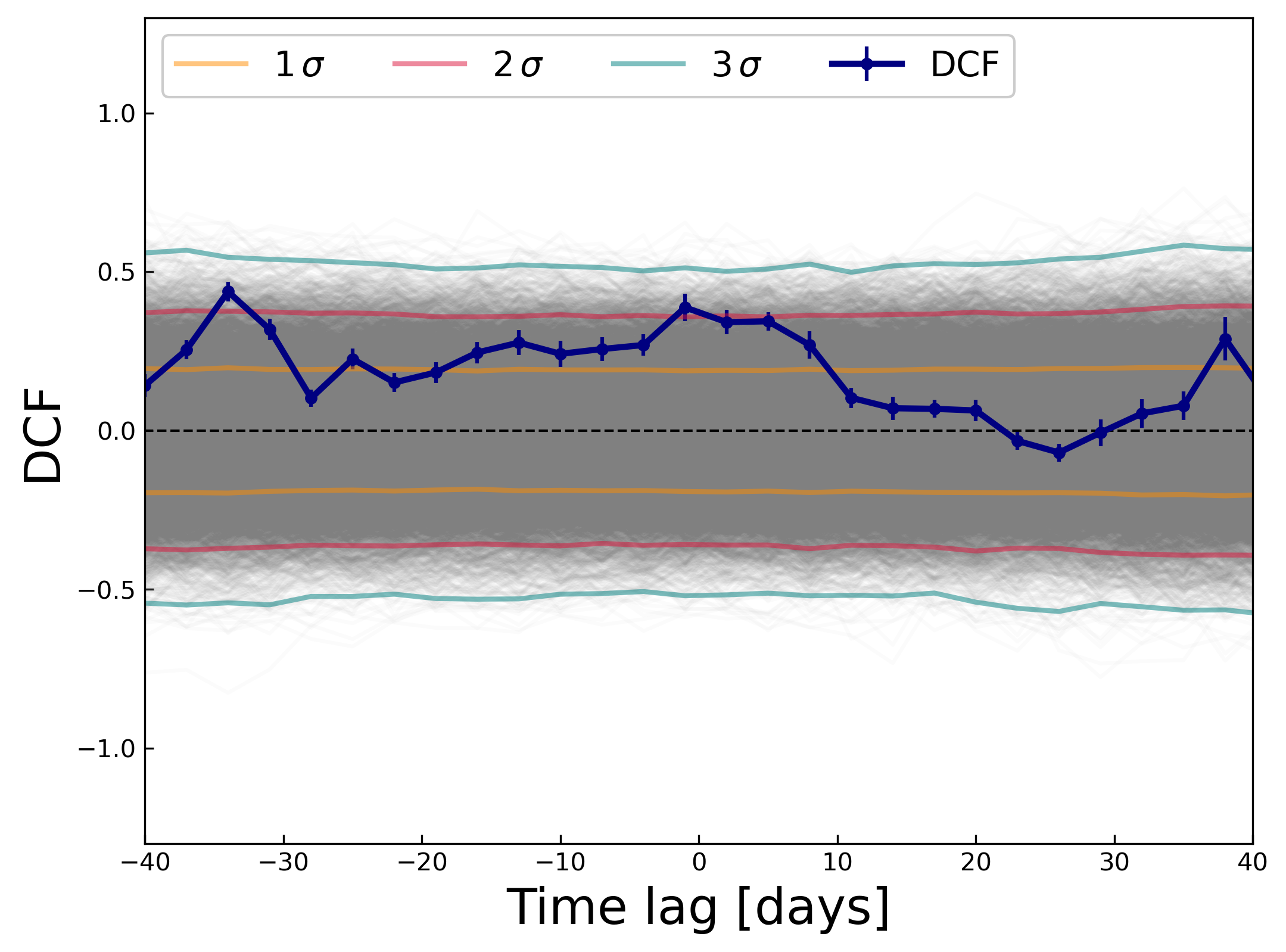}
         \caption{}
         \label{fig:Fermi_HE_vs_optical}
     \vspace{10pt}
     \end{subfigure}
     \hspace{15pt}
     \begin{subfigure}[t]{0.45\textwidth}
         \centering
         \includegraphics[width=\textwidth]{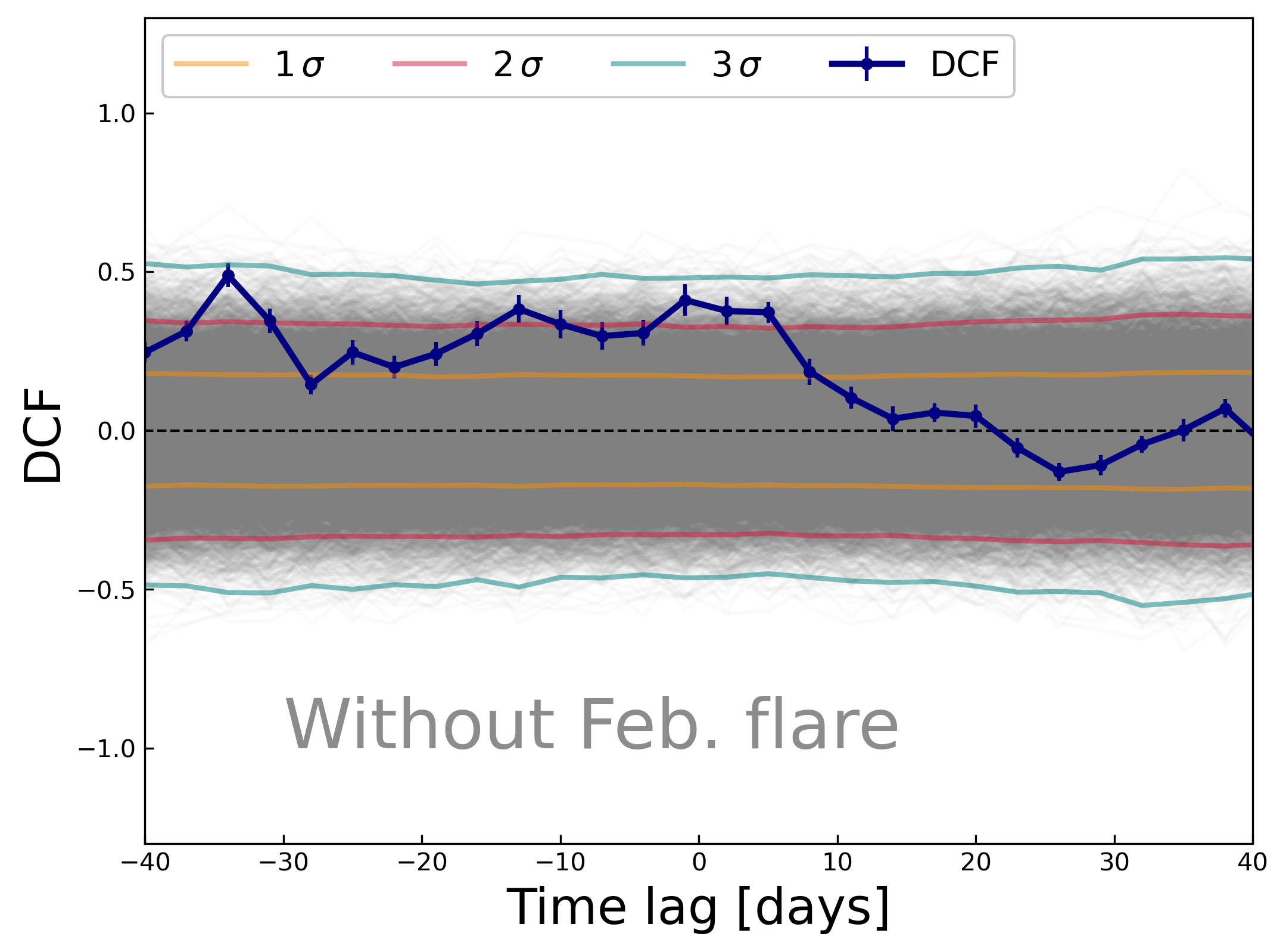}
         \caption{}
         \label{fig:Fermi_HE_vs_optical_noflare}
     \end{subfigure}

     \begin{subfigure}[t]{0.45\textwidth}         
     \centering
         \includegraphics[width=\textwidth]{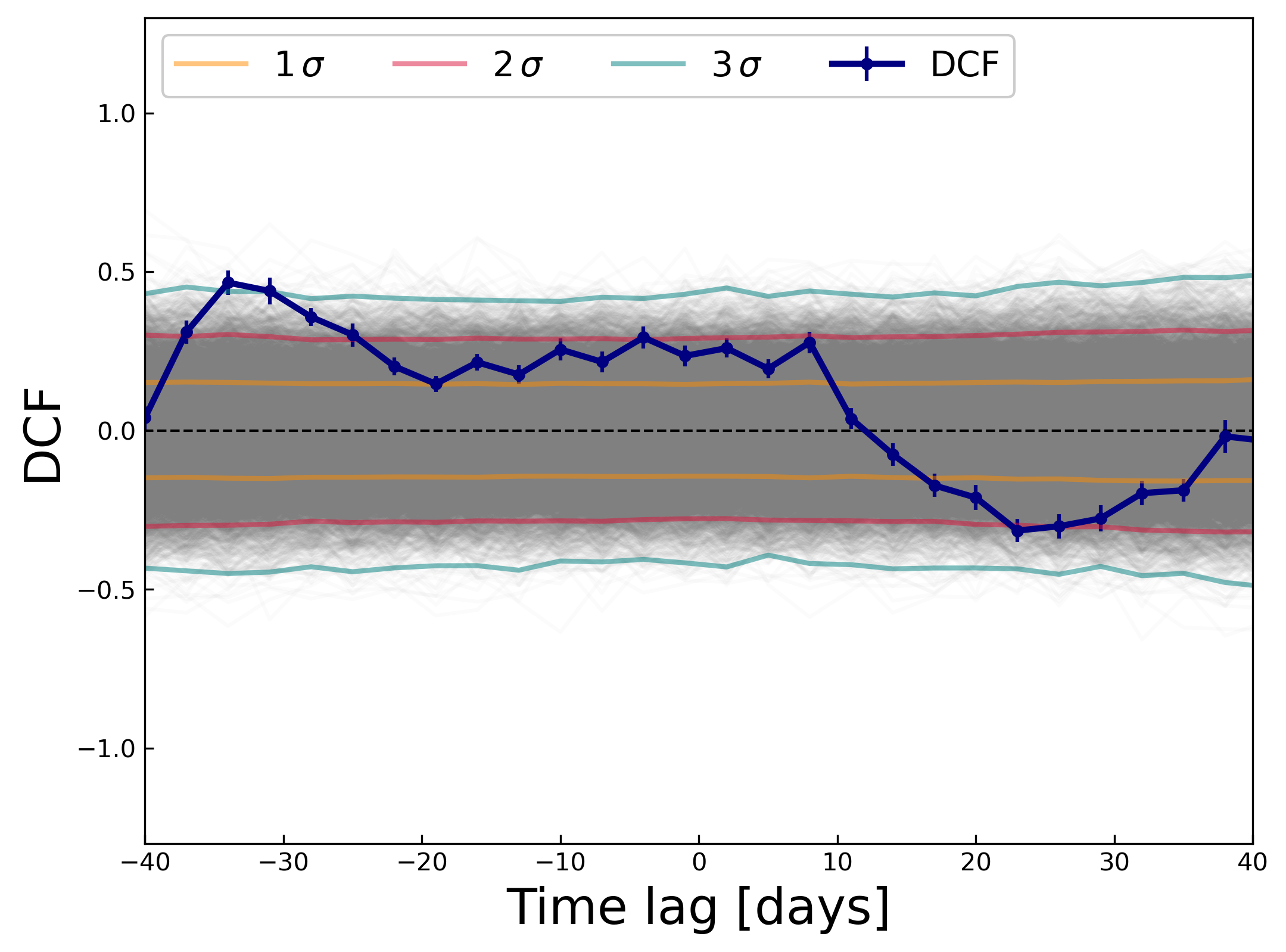}
         \caption{}
         \label{fig:Fermi_LE_vs_optical}
     \end{subfigure}
     \hspace{15pt}
     \begin{subfigure}[t]{0.45\textwidth}
         \centering
         \includegraphics[width=\textwidth]{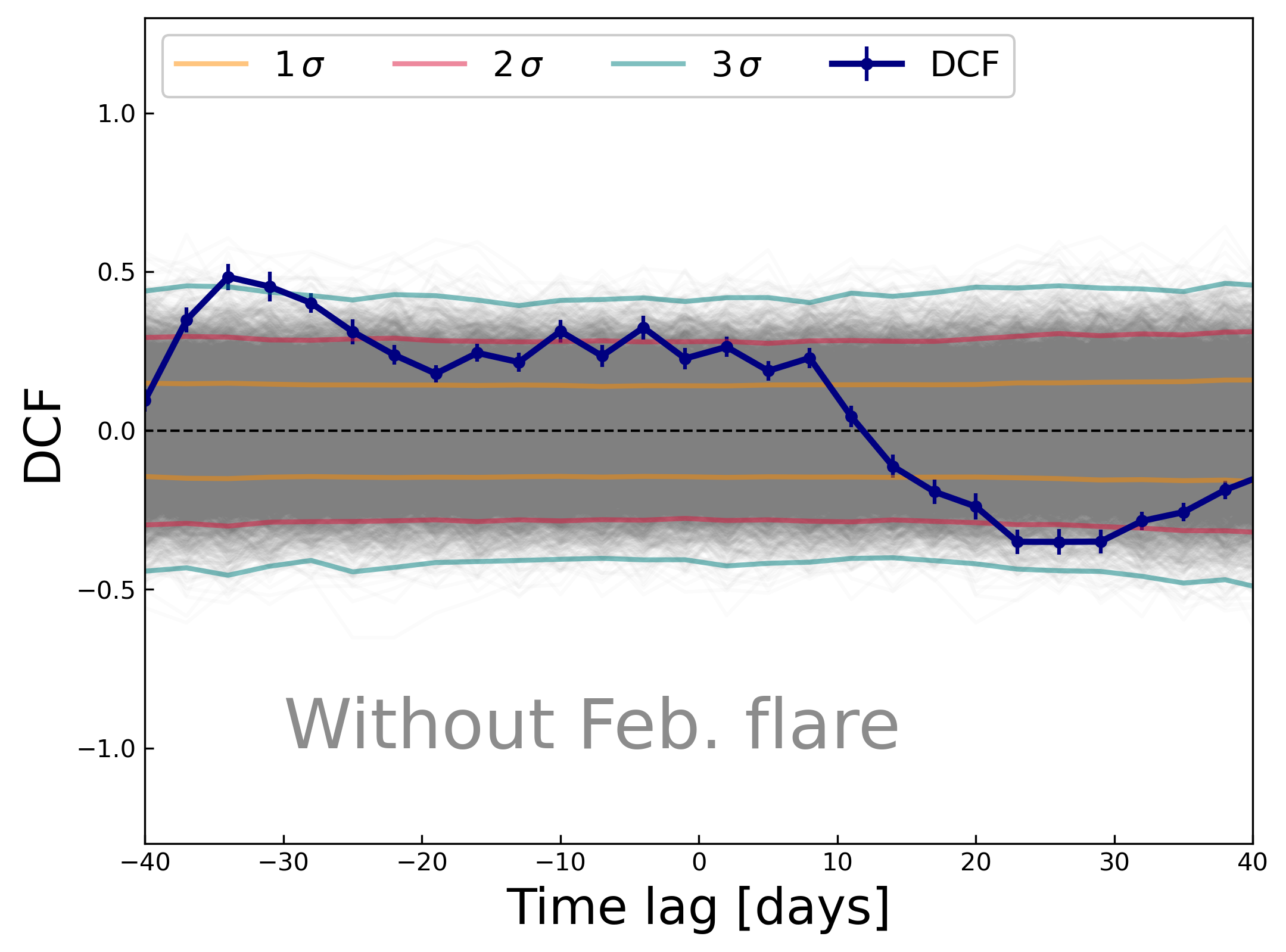}
         \caption{}
         \label{fig:Fermi_LE_vs_optical_noflare}
     \end{subfigure}
    \caption{Discrete correlation function computed between the two energy ranges provided by \textit{Fermi}-LAT and the R-band by GASP-WEBT without the big flare in February using a binning of 3 days. It is computed for a range of time lags between -40 to +40 days. The 1\,$\sigma$, 2\,$\sigma$, and 3\,$\sigma$ confidence levels obtained by simulations are shown by the yellow, red, and green lines, respectively. (a) 3-300\,GeV versus R-band; (b) 3-300\,GeV versus R-band without the flare in February 2010; (c) 0.3-3\,GeV versus R-band; (d) 0.3-3\,GeV versus R-band without the flare in February 2010.}
        \label{fig:Fermi_vs_optical}
\end{figure*}

\end{appendix}

\end{document}